\documentclass[aps,prb,twocolumn,showpacs]{revtex4}
\usepackage{graphicx}
\usepackage{psfrag}

\begin{document}

\title{Theory of 2D superconductor with broken inversion symmetry}

\author{Ol'ga Dimitrova}
\affiliation{L.D.Landau Institute for Theoretical Physics, Moscow,
119334, Russia} \affiliation{The Abdus Salam International Center
for Theoretical Physics, Strada Costiera 11, 34100 Trieste, Italy}
\author{M. V. Feigel'man}
\affiliation{L.D.Landau Institute for Theoretical Physics, Moscow,
119334, Russia}

\date{\today}

\begin{abstract}
A detailed theory of a phase diagram of a 2D surface
superconductor in a parallel magnetic field is presented. A
spin-orbital interaction of the Rashba type is known \cite{BG} to
produce at a high magnetic field $h$ (and in the absence of
impurities)  an inhomogeneous superconductive phase similar to the
Larkin-Ovchinnikov-Fulde-Ferrel (LOFF) state with an order
parameter $\Delta({\bf r})\propto \cos({\bf Qr})$. We consider the
case of a strong Rashba interaction with the spin-orbital
splitting much larger than the superconductive gap $\Delta$, and
show that at low temperatures $T\leq 0.4 T_{c0}$ the LOFF-type
state is separated from the usual homogeneous state by a
first-order phase transition line. At higher temperatures  another
inhomogeneous state with $\Delta({\bf r}) \propto \exp(i{\bf Qr})$
intervenes between the uniform BCS state and the LOFF-like state
at $g\mu_B h \approx 1.5 T_{c0}$.  The modulation vector $Q$ in
both phases is of the order of $g\mu_B h/v_F$. The superfluid
density $n_s^{yy}$ vanishes in the region around the second-order
transition line between the BCS state and the new ``helical''
state. Non-magnetic impurities suppress both inhomogeneous states,
and eliminate them completely at $T_{c0}\tau \leq 0.11$. However,
once an account is made of the next-order term over the small
parameter $\alpha/v_F \ll 1$, a relatively long-wave helical
modulation with $Q \sim g\mu_B h\alpha/v_F^2$ is found to develop
from  the BCS state. This long-wave modulation is stable with
respect to disorder.   In addition, we predict that unusual vortex
defects with a continuous  core  exist near the phase boundary
between the helical and the LOFF-like states. In particular, in
the LOFF-like state these defects may carry a half-integer flux.

\end{abstract}


\maketitle

\section{Introduction}
\label{Introduction} There are experimental indications~\cite{WO3}
in favor of existence of superconductive states localized on a
surface of non-superconductive (or even insulating) bulk material.
Such a superconductive state should possess a number of unusual
features due to the absence of the symmetry ``up'' v/s ``down''
near the surface: the condensate wave-function is neither singlet
nor triplet, but a mixture of both~\cite{GR,Ed}; the Pauli
susceptibility is enhanced  at low temperatures~\cite{GR} (as
compared with the usual superconductors); the paramagnetic
breakdown of the superconductivity in a parallel magnetic field is
moved towards much higher field values due to a formation of an
inhomogeneous superconductive state~\cite{BG} similar to the one
predicted by Larkin-Ovchinnikov  and Fulde-Ferrel~\cite{LO,FF}
(LOFF) for a ferromagnetic superconductor. All these features
steam from the chiral subband splitting of the free electron
spectrum at the surface, due to the presence of the spin-orbital
Rashba term~\cite{rashba}; the magnitude of this splitting $\alpha
p_F$ is small compared to the Fermi energy but can be rather large
with respect to other energies in the problem. The line of
transition from normal to (any of) superconductive state $T_c(h)$
was determined in~[\onlinecite{BG}]; however, the nature of the
transition between the usual homogeneous (BCS) superconductive
state at low fields and the LOFF-like state at high fields was not
studied.

\begin{figure*}
\includegraphics[angle=0,width=5.7in]{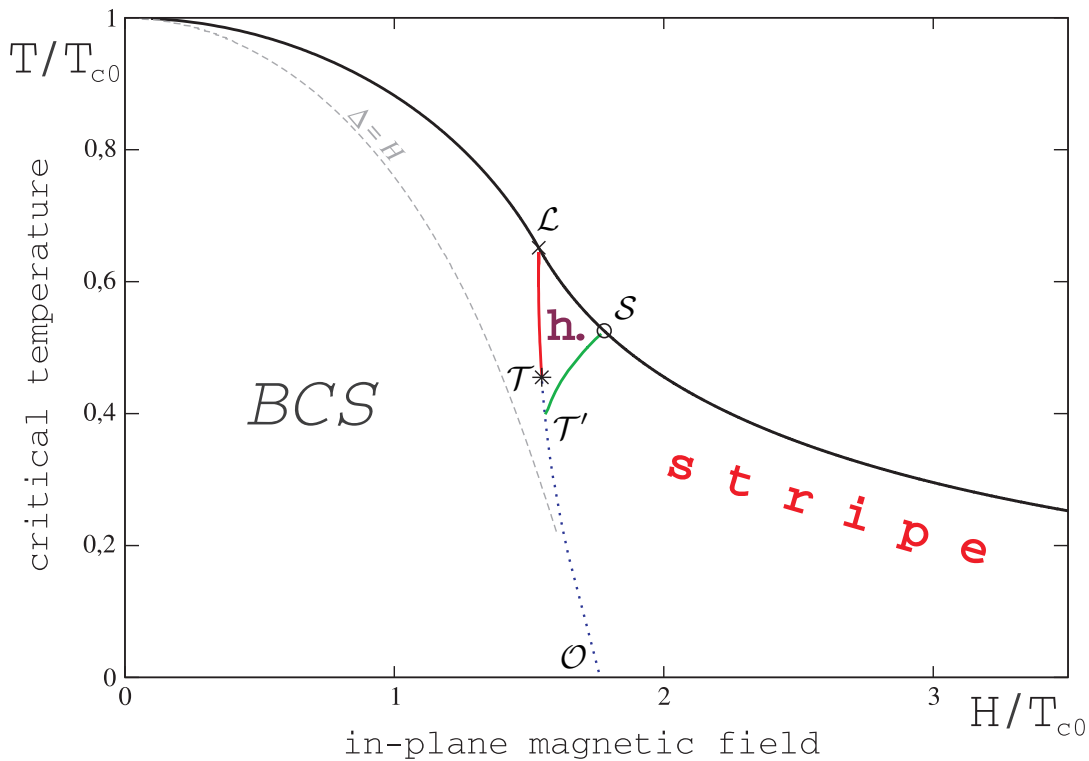}
\caption{\label{phd}  A phase diagram that shows: a
superconducting phase transition line $T_c(H)$ (solid) and two
second-order phase transition lines in the clean case, an
$\cal{L}\cal{T}$ line between the homogeneous (BCS) and the
helical (h.) state, and an $\cal{S}\cal{T'}$ line of stability of
the helical state. The dotted line going downwards from point
$\cal{T}$ to point ${\cal O}$ marks the absolute limit of
stability of the BCS state. The cross indicates the Lifshitz point
$\cal{L}$ and the circle indicates the symmetric point $\cal{S}$.
The line of transition into the gapless superconductivity
$H=\Delta$ is marked with a dashed line.}
\end{figure*}

In this paper we provide a detailed study of the phase diagram of
a surface superconductor in a parallel magnetic field $h$ (a brief
account of our results was published in Ref.~\onlinecite{DF2003}).
We show that at moderate values of $h \sim T_c/\mu_B$ the behavior
of this system is rather different from 2D LOFF model
of~[\onlinecite{rainer}]. Namely, we demonstrate the existence of
a {\it short-wavelength} helical state with an order parameter
$\Delta \propto \exp(i{\bf Qr})$ \, (where ${\bf Q} \perp {\bf
h}$\, and $Q \sim \mu_B h/v_F$) in a considerable part of the
phase diagram, which  is summarized in Fig.~\ref{phd}. The line
${\cal L}{\cal T}$ is the second-order transition line separating
the helical state from the homogeneous superconductor. Below the
${\cal T}$ point a direct first-order transition between the
homogeneous and the LOFF-like state takes place. The line ${\cal
TO}$ shown in Fig.~\ref{phd}  marks the border of metastability of
the BCS state; we expect that the actual first-order transition
line is (slightly) shifted towards lower $H$ values. The line
${\cal ST'}$ marks the second-order transition between the helical
and the LOFF-like state. The above results are valid within the
leading (the zeroth-order) approximation over the small parameter
$\alpha/v_F \ll 1$;  an account of the terms linear in $\alpha/v_F
\ll 1$ leads, in agreement with~[\onlinecite{Agterberg}], to the
transformation of the uniform BCS state into a {\it
long-wavelength helical} state with a wave vector $q \sim \alpha
H/v_F^2$  at the lowest magnetic fields. Therefore, the ${\cal
LT}$ line is actually a line of a sharp crossover (with a relative
width of the order of $\alpha/v_F$) between the long- and the
short- wavelength phases.

The rest of the paper is organized as follows. In Sec.~\ref{Model}
we introduce a model of a spin-orbital superconductor in a
parallel magnetic field, with a hierarchy of the energies
$\epsilon_F\gg \alpha p_F\gg T_c$. In Sec.~{\ref{SCtransition}} we
derive the Ginzburg-Landau functional for an inhomogeneous ground
state. On the $T_c(H)$ line we find two critical points: the
Lifshitz point $\cal{L}$ and the symmetric point $\cal{S}$, and
demonstrate the existence of a ``helical'' state with an order
parameter $\Delta({\bf r})\propto\exp(i{\bf Qr})$ and a {\it
large} $Q \sim H/v_F$. The point ${\cal S}$ on the $T_c(H)$ line
is special in the sense that there the order parameter symmetry is
enhanced to $U(2)$ from the usual $U(1)$. That leads to unusual
vortex textures discussed in Sec.~\ref{SU2}. In particular,
vortices with half-integer flux are predicted for the LOFF-like
state. In Sec.~\ref{PhaseDiagram} we derive the two stationary
conditions for the helical state, which determine the equilibrium
$\Delta$ and $Q$. The latter allows to establish the boundaries of
stability of the BCS and helical state: the Lifshitz line
terminating in the critical Landau point $\cal{T}$, and a line
starting in the symmetric point $\cal{S}$. We show that the
helical state and the parity-even (stripe) phase are separated by
two phase transitions of the second order and an intervening
additional superconducting phase. In Sec.~\ref{Current} we prove
that the electric current is zero in the ground state; then we
show that the supercurrent response to the vector potential
component $A_y = {\bf A}{\bf Q}/Q$ vanishes at the ${\cal LT}$
line, whereas within the helical state both components of the
superfluid density tensor $\hat{n}_s$ are of the same magnitude as
the superfluid density in the BCS state. In Sec.~\ref{WeakHelical}
we explore the influence upon the phase diagram of the terms
linear in $\alpha/v_F \ll 1$. We show that at low magnetic fields
the ground state is realized as a weakly helical state with zero
current. A special geometry is proposed when an oscillating
supercurrent in the ground state of the helical phase may flow. In
Sec.~\ref{Impurities} we study the effect of the non-magnetic
impurities on the phase diagram. We show that in the relatively
clean case the paramagnetic critical field is quickly suppressed
by disorder and the position of the Lifshitz point is shifted
towards higher magnetic field values. We find the critical
strength of disorder above which all short-wavelength
inhomogeneous states are eliminated from the phase diagram; in
terms of an elastic scattering time $\tau$ this condition reads $
\tau_c = 0.11 \hbar/T_{c0}$. At $\tau < \tau_c$ the only phase
which survives is the ``weakly helical'' state with $Q = 4\alpha
H/v_F^2$; in this regime the paramagnetic critical field starts to
{\it increase} with disorder, and at $\tau \ll \tau_c$ we find
$H_c \propto \tau^{-1/2}$. In Sec.~\ref{BKTtransition} we go
beyond the mean-field approximation and study the modifications of
the transition line $T_c(H)$ due to the
Berezinsky-Kosterlitz-Thouless vortex depairing transition. We
demonstrate that vortex fluctuations are strongly enhanced near
the points  ${\cal L}$ and ${\cal S}$, leading to local downward
deformations of the actual $T_{BKT}(H)$ line.

\section{Model of a spin-orbital superconductor}
\label{Model}

Near the surface of a crystal translational symmetry is reduced
and inversion symmetry is broken even if it is present in the
bulk. (The component of the electron momentum $\hat{\vec{p}}$
parallel to the interface is conserved because of the remaining 2D
translational symmetry.) As a result a transverse electrical field
appears near the surface. The electron spin couples to this
electric field due to the Rashba spin-orbit
interaction~\cite{rashba}
\begin{equation}
H_{so}=\alpha\left[\hat{\vec{\sigma}} \times
  \hat{\vec{p}}\right]\cdot\vec{n},
\label{Rashba1}
\end{equation}
where $\alpha>0$ is the spin-orbit coupling constant, $\vec{n}$ is
a unit vector perpendicular to the surface,
$\hat{\vec{\sigma}}=(\hat\sigma_x,\hat\sigma_y,\hat\sigma_z)$ are
the Pauli matrices. This interaction explicitly violates inversion
symmetry. The electron spin operator does not commute with the
Rashba term, thus the spin projection is not a good quantum
number. On the other hand, the chirality operator
$[\hat{\vec{\sigma}}\times \hat{\vec{e}}]\cdot\vec{n}$ commutes
with the Hamiltonian. Here $\hat{\vec{e}}=\hat{\vec{p}}/p$ is the
momentum direction operator with eigenvalues $\vec{e}_{\bf
p}=(\cos\varphi_{\bf p}, \sin\varphi_{\bf p})$, where
$\varphi_{\bf p}$ is the angle between the momentum of the
electron and the $x$-axis. The chirality operator eigenvalues
$\lambda=\pm 1$ together with the momentum constitute the quantum
numbers of the electron state $(\vec{p},\lambda)$. The Rashba term
(\ref{Rashba1}) preserves the Kramers degeneracy of the electron
states, thus the states $(\vec{p},\lambda)$ and
$(-\vec{p},\lambda)$ belong to the same energy.

\psfrag{L}[c][c][2.3][0]{\kern-1pt\lower0pt\hbox{${\cal L}$}}
\psfrag{S}[c][c][2.3][0]{\kern-1pt\lower2pt\hbox{${\cal S}$}}
\psfrag{T}[c][c][2.3][0]{\kern0pt\lower0pt\hbox{${\cal T}$}}
\psfrag{T1}[c][c][2.3][0]{\kern0pt\lower0pt\hbox{${\cal T'}$}}
\psfrag{O}[c][c][2.3][0]{\kern1pt\lower3pt\hbox{${\cal O}$}}

In this paper we consider the simplest model~\cite{GR} of a
surface superconductor: a BCS model for a two-dimensional metal
with the Rashba term (\ref{Rashba1}), in the limit $\alpha p_F\gg
T_c$. The Hamiltonian written in the coordinate representation
reads
\begin{eqnarray}\label{koord}
\hat{H}=\int \psi_{\alpha}^+(\vec{r})
\hat{h}
\psi_{\beta}(\vec{r})\,d^2\vec{r}-\frac{U}{2}\int
\psi_{\alpha}^+\psi_{\beta}^+\psi_{\beta}\psi_{\alpha}
\,d^2\vec{r}\,,
\end{eqnarray}
with the one-particle Hamiltonian operator
\begin{eqnarray}\label{koord1}
\hat{h}=\left
(\frac{\hat{P}^2}{2m}\delta_{\alpha\beta}
+\alpha\left [
\hat{\vec{\sigma}}_{\alpha\beta}\times
\hat{P}\right ]\cdot
\vec{n}-g\mu_B\vec{h}\cdot \hat{\vec{\sigma}}_{\alpha\beta}/2
\right),\,
\end{eqnarray}
where $m$ is the electron mass, $\alpha$, $\beta$ are the spin
indices and $\hat{P}=-i\vec{\nabla}-\frac{e}{c}\vec{A}(\vec{r})$
is the momentum operator in the presence of an infinitesimal
in-plane vector potential $\vec{A}=\vec{A}(\vec{r})$, $e<0$ is the
electron charge. We have included into the Hamiltonian the Zeeman
interaction with a uniform external magnetic field $\vec{h}$
parallel to the interface, assuming $\vec{h}$ to be in the
$x$-direction. The vector potential of such a field can be chosen
to have only the $z$-component, therefore it decouples from the 2D
kinetic energy term. $\mu_B$ is the Bohr magneton and $g$ is the
Lande factor. Hereafter we use a notation $H=g\mu_Bh/2$.

The electron operator can be expanded in the basis of plane waves:
$\hat{\psi}_\alpha(\vec{r})= \sum_{{\bf p},\lambda}\ e^{i{\bf
p}\vec{r}}a_{\alpha {\bf p}}$, and the one-particle part of the
Hamiltonian (\ref{koord}) in the momentum representation can be
written as a sum of $\hat{H}_0$ and $\hat{H}_{em}$:
\begin{eqnarray}\label{momrep}
&&\hat{H}_0=\sum_{{\bf{p}}}a_{\alpha {\bf p}}^{+}\left
( \frac{{\bf p}^2}{2m}\hat{1}+\alpha\left[
\hat{\vec{\sigma}}_{\alpha\beta} \times {\bf p}\right]
\cdot\vec{n} - \vec{H}\cdot
\hat{\vec{\sigma}}_{\alpha\beta} \right )a_{\beta {\bf p}},
\nonumber \\
&&\hat{H}_{em}=\sum_{{\bf{p}}}a_{\alpha {\bf p}}^{+}\left
(-\frac{1}{c}\,\,\hat{\vec{j}}\vec{A}\right )a_{\beta {\bf p}}.
\nonumber \\
\end{eqnarray}
Here the current operator is
\begin{equation}\label{spincurrent}
\hat{\vec{j}}=-e(\hat{\vec{p}}/m-\alpha [\hat{\vec{\sigma}} \times
\vec{n}])-\frac{e^2}{2mc}\vec{A}.
\end{equation}

\begin{figure}
\includegraphics[angle=0,width=0.19\textwidth]{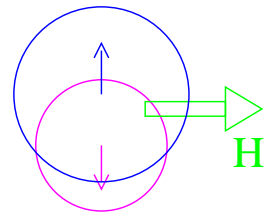}
\caption{\label{Fermi} When an external magnetic field $H\ll\alpha
p_F$ is applied in the $x$-direction, the two Fermi circles
corresponding to the different chiralities $\lambda=\pm 1$ are
shifted in opposite $y$-directions by a small momentum $Q=\pm
H/v_F$. The circle corresponding to one of the chiralities is
larger than the other by the relative amount of $\alpha/v_F$.}
\end{figure}

The Hamiltonian $\hat{H}_{0}$ can be diagonalized by the
transformation
$a_{\alpha {\bf p}}=
\eta_{\lambda\alpha}({\bf p})\hat{a}_{\lambda{\bf p}}$
with the
two-component spinor
\begin{equation}\label{eigenf}
\eta_{\lambda}({\bf p})={1\over\sqrt{2}}
\left(\begin{array}{c}
\displaystyle 1 \\ \displaystyle
i\lambda\exp(i\varphi_{\bf p}(H))\end{array}\right),
\end{equation}
where
\begin{equation}\label{phipH}
\varphi_{\bf p}(H)=\arcsin \frac{\alpha p_{y}-H}
{\sqrt{(\alpha p)^2-2\alpha p_{y}H+H^2}}.
\end{equation}
The eigenvalues of the Hamiltonian (\ref{momrep}), corresponding
to the chiralities $\lambda=\pm 1$, are
\begin{equation}\label{eigenv}
\epsilon_\lambda({\bf p})=p^2/2m -\lambda\sqrt{(\alpha p)^2-2\alpha
p_{y}H+H^2},
\end{equation}
thus at $H=0$  the equal-momentum electron states are split in
energy by $2\alpha p_F$. The two Fermi circles, corresponding to
the different chiralities, have Fermi-momenta $p^{\pm}_F=
\sqrt{2m\mu+ m^2\alpha^2} \pm m\alpha$, where $\mu\gg m\alpha^2$
is the chemical potential. The density of states on the two Fermi
circles is almost the same, $\nu_{\pm}=\frac{m}{2\pi}\left(1\pm
\frac{\alpha}{v_F}\right)$. In the main part of the paper we
neglect the difference $\nu_+-\nu_-$. The effects related to
$\nu_+ \neq \nu_-$ will be considered in Sec.~\ref{WeakHelical}.
When an external magnetic field is applied, these two Fermi
circles are displaced in opposite $y$-directions by a momentum
$Q=\pm H/v_F$, as shown in Fig.~\ref{Fermi}.

The two-particle pairing interaction in Hamiltonian (\ref{koord})
in the momentum representation reads
\begin{equation}
-\frac{U}{2}\sum_{{\mathbf p}{\mathbf p'}{\mathbf q}}
a_{\alpha{\bf p}+{\mathbf q}/2}^{+}
a_{\beta-{\mathbf p}+{\mathbf q}/2}^{+}
a_{\beta-{\mathbf p'}+{\mathbf q}/2}
a_{\alpha{\mathbf p'}+{\mathbf q}/2},
\end{equation}
and can be simplified in the chiral basis (\ref{eigenf}), assuming
$H\ll\alpha p_F\ll\mu$. In the long-wavelength limit $q\ll p_F$ it
can be factorized as
\begin{equation}\label{Hint}
\hat{H}_{int}=
-\frac{U}{4}\sum_{{\bf q}}\hat{A}^+({\bf q})\hat{A}({\bf q}),
\end{equation}
where the pair annihilation operator
\begin{equation}\label{annihilation}
\hat{A}({\bf q})=\sum_{{\bf p},\lambda}\lambda
e^{i\varphi_{\bf p}}
a_{\lambda-{\bf p}+{\bf q}/2}a_{\lambda{\bf p}+{\bf q}/2 }.
\end{equation}
Here $\lambda e^{i\varphi_{\bf p}}$ is the wave function of
the Cooper pair in the chiral basis and it changes sign under the
substitution $-{\bf p}$ for ${\bf p}$.

To calculate the thermodynamic potential $\Omega=-T\ln{Z}$, we
employ the imaginary-time functional integration technique with
the Grassmanian electron fields $a_{\lambda {\bf p}},
\bar{a}_{\lambda {\bf p}} $ and introduce an auxiliary complex
field $\Delta({\bf r},\tau)$ to decouple the pairing term
$H_{int}$, cf.~[\onlinecite{Popov}]. The resulting effective
Lagrangian is
\begin{eqnarray}
&&L[a,\bar{a},\Delta,\Delta^*]=
\sum_{{\bf p},\lambda}\bar{a}_{\lambda{\bf p}}\left
(-\partial_{\tau}-\epsilon_\lambda({\bf p})\right)a_{\lambda{\bf p}}+
\nonumber \\
&&+\sum_{\bf
q} \Big[-\frac{|\Delta_{\bf q}|^2}{U}+\frac{1}{2}\sum_{{\bf p},\lambda}
\big(\Delta_{\bf q}\lambda e^{-i\varphi_{\bf p}}
\bar{a}_{\lambda,{\bf p}+{\bf q}/2}
\bar{a}_{\lambda,-{\bf p}+{\bf q}/2} \nonumber \\
&&+\Delta^*_{\bf q}\lambda e^{i\varphi_{\bf p}}
 a_{\lambda,-{\bf p}+{\bf q}/2}
 a_{\lambda, {\bf p}+{\bf q}/2}\Big].
\label{Lhub-str}
\end{eqnarray}
The thermodynamic potential $\Omega=-T\ln{Z}$ describes a system
in equilibrium, where $Z$ is the grand partition function. We
express $\Omega$ as a zero-field limit of a generating functional
$\Omega=\Omega[\eta,\bar{\eta}]\big|_{\eta\to 0}$:
\begin{eqnarray}\label{funct-int}
\exp\left(\!-\frac{\Omega[\eta,\bar{\eta}]}{T}\right)\!=
\!\!\!\int\!{\cal D}\Delta{\cal D}\Delta^*
\exp\left(-\frac{\Omega[\eta,\bar{\eta},\Delta,\Delta^*]}{T}\right)\!
=\nonumber \\
=\!\int\!{\cal D}a{\cal D}\bar{a}{\cal D}\Delta{\cal D}\Delta^*
\exp\Big (\int_0^{1/T}\big[ L[a,\bar{a},\Delta,\Delta^*]+
\nonumber \\
+\sum_{{\bf p}} (\bar{\eta}(\tau {\bf p}) a(\tau
{\bf p}) +\bar{a}(\tau {\bf p})\eta(\tau {\bf p}) )\big ]d\tau\Big).
\nonumber \\
\end{eqnarray}

Below we will work within the mean-field approximation, which is
controlled by the smallness of the Ginzburg number ${\rm Gi}$.
However, for a clean 2D superconductor ${\rm Gi} \sim T_c/E_F$ may
be non-negligible (we discuss the fluctuational effects in the end
of this paper). The mean-field approximation is equivalent to the
saddle-point approximation for the functional integral over
$\Delta$ and $\Delta^*$  defined in the first line in
Eq.~(\ref{funct-int}). In other terms,  we will study  the minima
of the functional $\Omega[\Delta,\Delta^*]$, which comes about
after integration over the Grassmanian fields in the functional
integral defined in the second line of Eq.~(\ref{funct-int}),
\begin{equation}\label{Extrem}
\frac{\delta \Omega[\Delta,\Delta^*]}{\delta \Delta(\tau{\bf
r})}=0.
\end{equation}
To evaluate the thermodynamic potential $\Omega[\Delta,\Delta^*]$,
we will use the Green's function method. The electron Green's
function is defined as a variational derivative of the generating
functional:
\begin{equation}\label{GreenFuncDef}
G(\tau {\bf r},\tau' {\bf r}')=\frac{\delta \Omega[\eta,\bar{\eta}]}
{\delta
\bar{\eta}(\tau{\bf r})\delta \eta(\tau'{\bf r}')}\Big|_{\eta\to 0}.
\end{equation}

In the next Section we determine the line of the superconducting
transition $T_c(H)$, and locate two special points on this line,
${\cal L}$ and ${\cal T}$, which designate the boundaries of
different superconductive states.

\section{Superconducting phase transition}
\label{SCtransition}

Near the phase transition from a normal metal to a superconductor
the order parameter $\Delta(\vec{r})$ is small. Therefore the
thermodynamic potential $\Omega$ may be expanded in powers of
$\Delta(\vec{r})$ and its gradients. This is known as the
Ginzburg-Landau functional. It has been shown by Barzykin and
Gor'kov~\cite{BG}, that the ground state can be inhomogeneous in
the direction perpendicular to the magnetic field. We consider the
order parameter as a superposition of a finite number of
harmonics:
\begin{equation}\label{DeltaHarmExp}
\Delta(\vec{r})= \sum_{i}\Delta_{{\bf
Q}_i}(\vec{r})\exp\left(i{\bf Q}_i\vec{r}\right),
\end{equation}
where $\Delta_{{\bf Q}_i}(\vec{r})$ are slowly varying envelope
functions. The corresponding Ginzburg-Landau functional is
\begin{eqnarray}\label{Ginzburg-Landau}
&& \Omega_{sn}=
\int
\Big[\sum_i\alpha_{Q_iQ_i}|\Delta_{{\bf Q}_i}(\vec{r})|^2+ \nonumber \\
&& +\sum_{ijkl}\beta_{Q_iQ_jQ_kQ_l}
\Delta_{{\bf Q}_i}(\vec{r})\Delta^*_{{\bf Q}_j}(\vec{r})
\Delta_{{\bf Q}_k}(\vec{r})\Delta^*_{{\bf Q}_l}(\vec{r})\times
\nonumber \\
&&\qquad\times\delta_{{\bf Q}_i+{\bf Q}_k-{\bf Q}_j-{\bf Q}_l}+
\nonumber \\
&&+\sum_ic_{Q_iQ_i}^x\left|\left(-i\hbar\frac{\partial}{\partial x}-
\frac{2e}{c}A_x(\vec{r})\right)\Delta_{{\bf Q}_i}(\vec{r})\right|^2+
\nonumber \\
&&+\sum_ic_{Q_iQ_i}^y\left|\left(-i\hbar\frac{\partial}{\partial y}-
\frac{2e}{c}A_y(\vec{r})\right)\Delta_{{\bf Q}_i}(\vec{r})\right|^2
\Big]
d^2\vec{r}.\nonumber \\
\end{eqnarray}

\begin{figure}
\includegraphics[angle=0,width=1.9in]{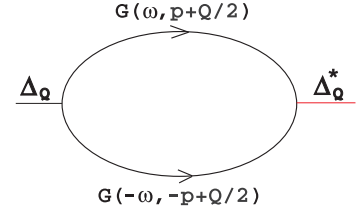}
\caption{\label{Cloop}
A Cooper loop with transferred momentum $Q$.}
\end{figure}

The coefficient $\alpha_{QQ}$ is given by the Cooper loop diagram
with transferred momentum $Q$. The coefficients $\beta_{QQQQ}$ and
$\beta_{QQ-Q-Q}$ are determined by four-Green's-function loop
integrals:
\begin{equation}\label{alphacoefficient}
\alpha_{QQ}=\frac{1}{U}-\frac{T}{2}\sum_{\omega,{\bf p},\lambda}
G_\lambda \left(\omega,{\bf p}+{\bf Q}/2
\right) G_\lambda\left(-\omega,-{\bf p}+{\bf Q}/2\right),
\end{equation}
\begin{eqnarray}\label{betacoefficients}
&&\beta_{QQQQ}=T\sum_{\omega,{\bf p},\lambda}G_\lambda^2
\left(\omega,{\bf p}+{\bf Q}/2\right) G^2_\lambda
\left(-\omega,-{\bf p}+{\bf Q}/2\right),\nonumber \\
&&\beta_{QQ-Q-Q}=T\sum_{\omega,{\bf p},\lambda}G_\lambda^2
\left(\omega,{\bf p}+{\bf Q}/2\right) \times \nonumber \\
&&\times G_\lambda
\left(-\omega,-{\bf p}+{\bf Q}/2\right)G_\lambda
\left(-\omega,-{\bf p}+3{\bf Q}/2\right),
\end{eqnarray}
\begin{eqnarray}\label{determine_c}
c_{QQ}^{\mu}=
\frac{1}{2}\frac{\partial^2}{\partial q_{\mu}^2}
\alpha(Q\vec{e}_y+q_{\mu}\vec{e}_{\mu}).
\end{eqnarray}
Here the normal state Green's function in an external
in-plane magnetic field $H\ll\alpha p_F$ is
\begin{equation}\label{normal}
G_\lambda\left(\omega,{\bf p}\right)=
\frac{1}{i\omega-\xi_\lambda({\bf p})-\lambda H\sin{\varphi_{\bf p}}},
\end{equation}
where $\omega=2\pi (n+{{1}\over{2}})T$ is the Matsubara frequency,
$n$ is an integer,
and the quasiparticle dispersion
\begin{eqnarray}\label{quasiparticle dispersion}
\xi_\lambda({\bf p})=p^2/2m-\lambda\alpha p_F-\mu
\end{eqnarray}
is assumed to be small compared to $\alpha p_F$. The integrals
over the momenta in (\ref{alphacoefficient}) and
(\ref{betacoefficients}) are calculated in the semiclassical
approximation:
\begin{equation}\label{quasiclassical}
\int\frac{d^2{\bf p}}{(2\pi)^2}=\nu_\lambda(\epsilon_F)
\int_{-\infty}^{\infty} d\xi_\lambda\int_{0}^{2\pi}
\frac{d\varphi}{2\pi}.
\end{equation}
In this Section we will calculate all diagrams in the zeroth order
over the parameter $\alpha/v_F \ll 1$ (i. e. we approximate
$\nu_\lambda(\epsilon_F)\approx \nu(\epsilon_F)$); the effect of
the  terms of the order of $O(\alpha/v_F)$ will be discussed in
Sec.~\ref{WeakHelical} below.

Integrating over the momenta $p$ in the Ginzburg-Landau functional
coefficients gives
\begin{equation}\label{alpha(T,H)}
\alpha_{QQ}=\frac{1}{U}-\pi \nu(\epsilon_F)T\sum_{\omega>0,\lambda}
\frac{1}{\sqrt{\omega^2+H_{\lambda}^2}},
\end{equation}
\begin{eqnarray}\label{B_1,2(H)}
\beta_{QQQQ}&=&\frac{\nu(\epsilon_{F})}{4}\pi T
\sum_{\omega>0,\lambda}
\frac{2\omega^2-H_{\lambda}^2}{(\omega^2+H_{\lambda}^2)^{5/2}},
\nonumber \\
\beta_{QQ-Q-Q}&=&\frac{\nu(\epsilon_{F})}{2}\pi T
\sum_{\omega>0,\lambda}
\frac{(2\omega^2+H_{\lambda}^2)}
{\omega^2(\omega^2+H_{\lambda}^2)^{3/2}}\frac{H_{\lambda}}{v_F Q},
\nonumber \\
\end{eqnarray}
\begin{eqnarray}\label{cxcy}
c_{QQ}^x=c_{-Q-Q}^x=
-\frac{1}{2}\left(\frac{v_F}{2}\right)^2\pi \nu(\epsilon_F)
T\sum_{\omega,\lambda}\frac{1}{\left(\omega^2+
H_{\lambda}^2\right)^{3/2}},
\nonumber \\
c_{QQ}^y=c_{-Q-Q}^y=
-\frac{1}{2}\left(\frac{v_F}{2}\right)^2\pi \nu(\epsilon_F)
T\sum_{\omega,\lambda}\frac{2H_{\lambda}^2-\omega^2}{\left(\omega^2+
H_{\lambda}^2\right)^{5/2}},
\nonumber \\
\end{eqnarray}
where
\begin{eqnarray}\label{Hlambda}
H_{\lambda}=\lambda H+v_{F}Q/2.
\end{eqnarray}
Note, that in the Ginzburg-Landau functional
(\ref{Ginzburg-Landau}) the coefficients
$\beta_{QQQQ}=\beta_{-Q-Q-Q-Q}$ correspond to the terms
$|\Delta_Q|^4$ and $|\Delta_{-Q}|^4$; and the coefficient
corresponding to the term $|\Delta_Q|^2|\Delta_{-Q}|^2$ is a sum
of the four equal coefficients
$\beta_{QQ-Q-Q}=\beta_{Q-Q-QQ}=\beta_{-Q-QQQ}=\beta_{-QQQ-Q}$.

\begin{figure}
\includegraphics[angle=0,width=3.3in]{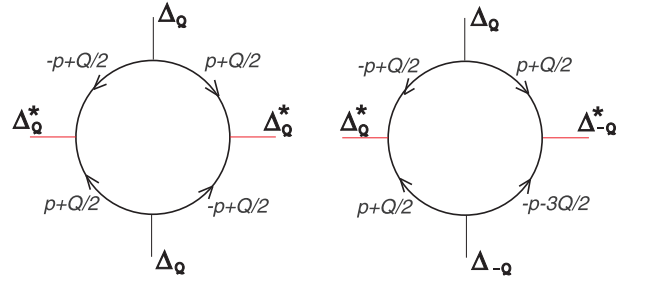}
\caption{\label{4order} Diagrams corresponding to the
terms of the fourth order in $\Delta$ in the Ginzburg-Landau
expansion: $\beta_{QQQQ}|\Delta_Q|^4$ and
$\beta_{QQ-Q-Q}|\Delta_Q|^2|\Delta_{-Q}|^2$.}
\end{figure}

The condition $\alpha_{QQ}=0$ determines the second-order
transition line (if $\beta_{QQQQ}>0$) between the normal metal and
the superconductor:
\begin{equation}\label{T_c(H)}
\frac{1}{U}=\nu(\epsilon_F)T\max_{Q} \sum_{\omega>0,\lambda}
\frac{\pi}{\sqrt{\omega^2+(\lambda H+v_{F}Q/2)^2}}.
\end{equation}

\psfrag{L}[c][c][3.5][0]{\kern-4.2pt\lower-1pt\hbox{${\cal L}$}}
\psfrag{S}[c][c][3.5][0]{\kern-5pt\lower1pt\hbox{${\cal S}$}}
\begin{figure}
\includegraphics[angle=0,width=3.4in]{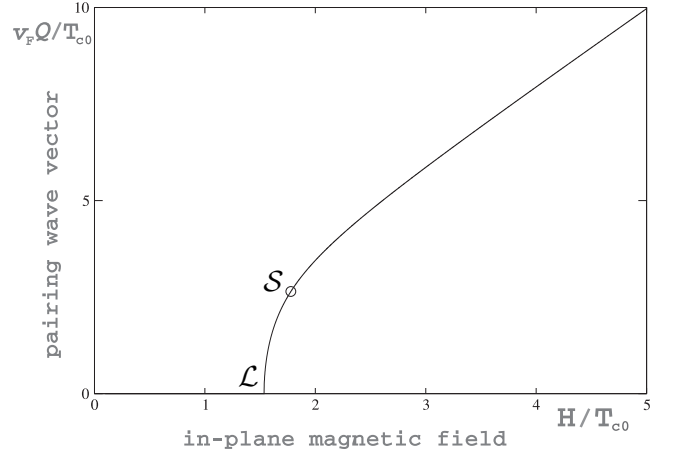}
\caption{\label{qh} Cooper pairing wave vector $Q.$ The circle
indicates the symmetric point $S$.}
\end{figure}
Equation (\ref{T_c(H)}) determines the shape of the phase
transition line $T_c(H)$ between the normal and superconducting
states. Depending upon $H$, the maximum over $Q$ in the r.h.s. of
Eq.~(\ref{T_c(H)}) is attained either at $Q=0$ or on nonzero
$Q=\pm |Q|$. The position of the $T_c(H)$ line found via numerical
solution of Eq.~(\ref{T_c(H)}) is shown in Fig.~\ref{phd}, where
both temperature $T_c$ and in-plane magnetic field $H$ are
normalized by the critical temperature at zero magnetic field:
$T_{c0}= 2\omega_D\exp(-1/\nu U+\gamma)/\pi$, where $\gamma=0.577$
is the Euler constant. The line $T_c(H)$ recovers two asymptotics
found in~[\onlinecite{BG}]:
\begin{eqnarray}
&&\log\frac{T_c(H)}{T_{c0}}=-\frac{7\zeta(3)H^2}{8\pi^2T_{c0}^2}
\qquad\mbox{ in the limit } H/T_{c0}\to 0,\nonumber \\
&&\frac{T_c(H)}{T_{c0}}=\frac{\pi T_{c0}}{2 e^\gamma H}
\qquad\mbox{ in the limit } H/T_{c0}\to \infty.
\end{eqnarray}

Whereas at low $H$ the superconductive solution is uniform, $Q=0$,
in the high field limit one finds~\cite{BG} $Q = 2H/v_F$. The
Lifshitz point ${\cal L}$ separates $Q=0$ and $Q\neq 0$ solutions
on the $T_c(H)$ line and is the end of the second-order phase
transition line between the two superconducting phases. In order
to determine the position of the ${\cal L}$ point, we note that
$\alpha_{QQ}$, cf. Eq.~(\ref{alpha(T,H)}), is symmetric under the
change $-Q \to Q$, and thus  it has always an extremum at $Q=0.$
Therefore  the position of the Lifshitz point ${\cal L}$ should
satisfy the equation
$$\left.\frac{\partial^2\alpha_{QQ}}{\partial Q^2}\right|_{Q=0}=0.$$
Numerical solution of the above equation gives the value
$(H_L,T_L)=(1.536,0.651) T_{c0}$, with the ratio
$H_{L}/T_{L}\approx 2.36$.

Fig.~\ref{qh} shows the Cooper pairing wave vector $Q$ on the
$T_c(H)$ line as a function of the in-plane magnetic field $H$.
Near the ${\cal L}$ point the wave-vector $Q$ contains a
square-root singularity $v_F Q(H)\sim \sqrt{H^2-H_L^2}$, typical
for the behavior of the order parameter near a second-order
transition. In the high-field limit $H/T_c(0) \to \infty$ the
behavior of the wave-vector $Q$ is given by the asymptotic
expression
\begin{equation}
v_F Q=2H-\frac{\pi^4 T_c^4(0)}{7\zeta(3)e^{2\gamma}H^3}.
\end{equation}
Note, that $Q=2H/v_F$ is the momentum shift of the two
$\lambda=\pm 1$ Fermi circles in a parallel magnetic field.

Near the $T_c(H)$ line the coefficient $\alpha$ can be approximated as
\begin{equation}
\alpha(T,H)\approx \nu(\epsilon_F)
\frac{T-T_c(H)}{2T} \sum_{\lambda} Y(T_c(H),H_\lambda),
\end{equation}
where $Y(T,\Delta)=
1-T\sum_{n}\frac{\Delta^2\pi}{(\omega_n^2+\Delta^2)^{3/2}}
=\frac{1}{4T}\int d\xi \mbox{
sech}^2\frac{\sqrt{\xi^2+\Delta^2}}{2T}$ is the Yosida function,
$\omega_n=\pi T (2n+1)$ is the Matsubara frequency. At $H > H_L$
an inhomogeneous superconductive phase is formed below the
$T_c(H)$ line. Eq.~(\ref{T_c(H)}) determines the absolute value of
the equilibrium wave vector $|Q|$ (the direction of ${\bf Q}$ is
perpendicular to ${\bf H}$), therefore two harmonics may
contribute to $\Delta(\vec{r})$ just below the $T_c(H)$ line:
$\Delta(y) = \Delta_+e^{iQy} + \Delta_- e^{-iQy}$. Below $T_c(H)$,
the density of the thermodynamic potential $\Omega$ is lower in
the superconductive state than in the normal one by the amount
\begin{eqnarray}\label{FF}
&&\Omega_{sn}=\alpha(T,H)|\Delta|^2 +\nonumber \\
&&\beta_{s}(T,H)|\Delta|^4 +
\beta_{a}(T,H)(|\Delta_+|^2 -|\Delta_-|^2)^2,
\end{eqnarray}
where $|\Delta|^2 =|\Delta_+|^2 + |\Delta_-|^2 $. Eq.~(\ref{FF})
was obtained from Eq.~(\ref{Ginzburg-Landau}) by
keeping of only two harmonics:
$|\Delta_+|^2=\Delta_{{\bf Q}}\Delta^*_{{\bf Q}}$,
$|\Delta_-|^2=\Delta_{-{\bf Q}}\Delta^*_{-{\bf Q}}$.
Comparing Eqs.~(\ref{Ginzburg-Landau}) and (\ref{FF})
gives
\begin{eqnarray}\label{betasa}
\beta_{s}(T,H)=\frac{1}{2}\beta_{QQQQ}+\beta_{QQ-Q-Q},\\ \nonumber
\beta_{a}(T,H)=\frac{1}{2}\beta_{QQQQ}-\beta_{QQ-Q-Q},
\end{eqnarray}
where $\beta_{QQQQ}$ and $\beta_{QQ-Q-Q}$ are the
four-Green's-function loop integrals (\ref{B_1,2(H)}). In the
symmetric point ${\cal S}$, where $\beta_{a}(T_c(H),H)=0$, the
free energy (\ref{FF}) depends upon $|\Delta|^2$ only, and thus is
invariant under $U(2)$ rotations of the order parameter spinor
$(\Delta_+,\Delta_-)$. The coordinates of the ${\cal S}$ point
are: $(H_S, T_S)=(1.779, 0.525)T_{c0}$, the corresponding wave
vector is $v_FQ_{S}=2.647T_{c}(0)$. At $ H < H_S$ we find
$\beta_{a} < 0$, and the free energy at $T < T_c(H)$ is minimized
by the choice of either $\Delta_+ = 0$ or  $\Delta_- = 0$, both
corresponding to the helical state. At $H > H_S$ $\beta_{a} > 0$
and the free energy minimum is achieved at $|\Delta_+| =
|\Delta_-|$, i. e. the LOFF-like phase with $\Delta(y) \propto
\cos(Qy)$ is the stable one at high field values.

\psfrag{L}[c][c][3.9][0]{\kern1.5pt\lower0pt\hbox{${\cal L}$}}
\psfrag{S}[c][c][3.9][0]{\kern-2pt\lower0pt\hbox{${\cal S}$}}
\begin{figure}
\includegraphics[angle=0,width=3.45in]{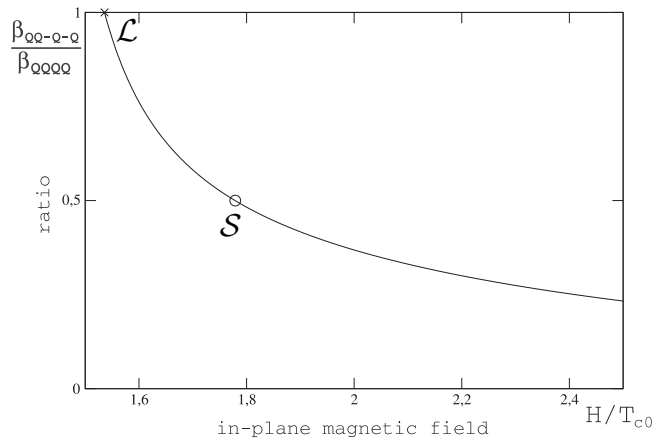}
\caption{\label{BB} The ratio of $\beta_{QQ-Q-Q}$ and
$\beta_{QQQQ}$ coefficients. The symmetric point ${\cal S}$,
marked on the figure with a circle, is defined as a point where
the ratio attains the value 1/2. The Lifshitz point ${\cal L}$ is
the point where the ratio equals 1, since $Q=0$ there.}
\end{figure}

\section{Unusual vortex solutions near the symmetric point ${\cal S}$.}
\label{SU2}
\subsection{General considerations:
extended $U(2)$ symmetry and vortices}

In this Section we discuss the peculiar properties of the
superconductive vortices, which appear due to the extended
symmetry of the order parameter near the symmetric point ${\cal
S}$ of the phase diagram. The free energy of the superconductor in
the vicinity of the symmetric point is given by the
Ginzburg-Landau functional
\begin{eqnarray}\label{segl}
\Omega_{sn} =\int d^2\vec{r}
\Big(c_i\left|\left(i\partial_{i}+\frac{2e}{c}A_i\right)
\Delta_+\right|^2
+\nonumber\\ \left.
+c_i\left|\left(i\partial_{i}+\frac{2e}{c}A_i\right)
\Delta_-\right|^2
+\beta_a(|\Delta_+|^2-|\Delta_-|^2)^2+ \right. \nonumber\\
+ \beta_s(\Delta_0^2- |\Delta_+|^2-|\Delta_-|^2)^2 -
\beta_s\Delta_0^4 \Big),
\end{eqnarray}
where $i=x,y$, and $\Delta_0^2=-\alpha/2\beta_s$ is the
equilibrium value of the order parameter. The fourth-order term in
the Ginzburg-Landau expansion~(\ref{FF}) can be divided into a
symmetric and an anisotropy part $\beta_{a}(T,H) (|\Delta_+|^2
-|\Delta_-|^2)^2$, and in the symmetric point the coefficient
$\beta_{a}(T_S,H_S)=0$. The coefficients $c_x$, $c_y$ are given by
the expressions (\ref{cxcy}), and their ratio in the $\cal{S}$
point is equal to $c_x/c_y=1.72$. Eq.~(\ref{segl}) can be written
in a rotationally symmetric form via an area-conserving
transformation by stretching and contracting the two coordinates
$x \to x(c_x/c_y)^{1/4}$ and $y \to y(c_y/c_x)^{1/4}$.

The minimum of the free energy~(\ref{segl}) in the symmetric point
(where $\beta_a=0$) is achieved in the homogeneous state under the
condition $|\Delta_+|^2+|\Delta_-|^2=\Delta^2$. In normalized
variables $z_1=\Delta_+/\Delta$ and $z_2=\Delta_-/\Delta$ the
order parameter spinor $(z_1,z_2)$ spans the sphere $S^3$:
$|z_1|^2+|z_2|^2=1$, and is equivalent to a four-component unit
vector $\vec{\cal{N}}$. This normalization allows us to write the
gradient part of the free energy~(\ref{segl}) as a non-linear
sigma-model
\begin{equation}\label{4N}
F_{\rm grad} =
\frac{\rho_s}{2} \int d^2\vec{r}(\partial_\mu\vec{\cal{N}})^2,
\end{equation}
where $\rho_s=\frac{|\alpha|}{\beta_s}\sqrt{c_xc_y}$ is defined
through coefficients of the Ginzburg-Landau
functional~(\ref{segl}).

Precisely at the symmetric point, the gradient functional
(\ref{4N}) governs the system's behavior at length scales $L$
larger than the order parameter correlation length $\xi(T)$. At
$\beta_a \neq 0$ its applicability is limited from the large
scales also. Namely, $L$ should be smaller than the
temperature-dependent ``anisotropy length''
\begin{equation}
L_{\rm an}(T) = \xi(T) \sqrt{\frac{\beta_s}{|\beta_a|}} \gg \xi(T),
\label{Lan}
\end{equation}
which is determined by comparison of
the gradient term and the anisotropy term in the full free energy
functional~(\ref{segl}).

On the left of the symmetric point ($\beta_a<0$)
the minimum of the energy is achieved at either $|z_1|=1$ or $|z_2|=1$,
leading to the degeneracy manifold of the order parameter,
 $S^1 \otimes Z_2$.
On the right of the symmetric point ($\beta_a>0$)
the degeneracy manifold of the order
parameter is $S^1\otimes S^1$.
In order for the gradient energy of a physical defect to be finite,
at large distances from the defect core the order parameter
should belong to the corresponding degeneracy manifold.
On the other hand, at relatively small distances
$ r \leq L_{\rm an} $, the whole
$S^3$ sphere is available for the order parameter configurations.

The two-dimensional $x$-space is topologically equivalent to a
sphere $S^2$ with a boundary at infinity $S^1$. A physical defect
is described as a mapping of a disk on the real plane $R^2$ with a
boundary $S^1$ (which encloses the defect) on the degeneracy
manifold of the order parameter $S^3$. Due to boundary conditions
at infinity (imposed by the anisotropy) the mapping
$S^2\rightarrow S^3$ is accompanied either by the mapping
$S^1\rightarrow S^1$  (at $\beta_a < 0$, i. e. on the left from
the ${\cal S}$ point), or by $S^1\rightarrow S^1 \otimes S^1 $
(at $\beta_a > 0$). Therefore  the topological defects are
determined by the nontrivial elements of the homotopy group
$\pi_2(S^3,S^1)=Z$ (on the left from the ${\cal S}$ point) or
$\pi_2(S^3,S^1\otimes S^1)=Z\otimes Z$ (on the right from the
${\cal S}$ point). Note that in the absence of any anisotropy
there would be no stable topological defects since $\pi_2(S^3)=0$,
i. e. any configuration of the order parameter could be
transformed into a homogeneous state. The general approach to the
classification of vortices with a nonsingular core by means of the
relative homotopy groups $\pi_2(R,\tilde{R})$, described above,
was first introduced by Mineev and
Volovik~[\onlinecite{MineevVolovik}]; a review of the approach can
be found in~[\onlinecite{Mineevreview}]. Some explicit solutions
for nonsingular vortices are presented in the review
[\onlinecite{Volovikreview}].

For an actual calculation it is convenient to employ the Hopf
projection, which splits the order parameter spinor $z\in S^3$,
parameterized here as
\begin{equation}\label{zEuler}
z=\left(\begin{array}{c} z_1\\ z_2\end{array} \right)= e^{i\chi}
\left(\begin{array}{c}e^{-i\varphi/2}\cos{\theta/2}\\
e^{i\varphi/2}\sin{\theta/2}\end{array} \right),
\end{equation}
into an $N=3$ - unit vector
$\vec{n}=z^\dagger\vec{\sigma}z$ $\in S^2$ sphere,
\begin{equation}\label{nEuler}
\vec{n}=
\left(\begin{array}{c}\sin{\theta}\cos{\varphi}\\
\sin{\theta}\sin{\varphi}\\
\cos{\theta}
\end{array} \right),
\end{equation}
parameterized through the Euler angles on the $S^2$ sphere, and a
total phase $\chi\in U(1)$, canonically conjugated to the electron
charge.

Each configuration $\vec{n}(x)$ defines a mapping of the
coordinate plane (equivalent to $S^2$) on a sphere $\vec{n}^2=1$,
i. e. the mapping $S^2\rightarrow S^2$. These mappings are
characterized by an integer ``topological charge''
\begin{equation}\label{charge}
{\cal Q}=
\frac{1}{4\pi}\int \limits_{R^2}\vec{n}
\left[\partial_x \vec{n}\times \partial_y\vec{n}\right]d^2\vec{r},
\label{Q1}
\end{equation}
which is related  with the circulation of the vector
\begin{eqnarray}\label{Aa}
{\cal{A}}_{\mu}=-i(z^\dagger\partial_{\mu}z-z\partial_{\mu}z^\dagger)=
(\partial_{\mu}\chi-\frac{1}{2}\partial_{\mu}\varphi\cos{\theta}).
\end{eqnarray}
Indeed the following identity can be proven:
\begin{equation}
{\cal Q} =
\frac{1}{2\pi}\oint\limits_{C_{\infty}}\vec{\cal{A}}\cdot\vec{d l} -
\frac{1}{2\pi}\oint\limits_{C_{0}}\vec{\cal{A}}\cdot\vec{d l}
\equiv \frac{\Phi}{\Phi_0} - C_0,
\label{indent}
\end{equation}
where $C_{\infty}$ is the closed  loop at infinity, and $C_0$ is
the infinitesimal closed loop just around the vortex singularity
point. The last equality in Eq.~(\ref{indent}) relates the
topological charge ${\cal Q}$ and the magnetic flux connected with
the vortex defect (in units of the superconductive flux quantum
$\Phi_0$).

Now  the gradient energy (\ref{4N}) can be represented as a sum of
the gradient energy of the $\vec{n}$-field and the kinetic term:
\begin{eqnarray}\label{includesigma}
F_{\rm grad} &=&
\frac{\rho_s}{2}\int \left[\frac{1}{4}(\partial_\mu \vec{n})^2+
{\cal{A}}_{\mu}^2 \right]   \ d^2\vec{r}.
\end{eqnarray}
Below we consider separately vortex solutions in the helical state
realized at $\beta_a < 0$ and in the LOFF-like state at $\beta_a > 0$.

\subsection{Non-singular vortices in the helical state.}

In the region on the left of the symmetric point ($\beta_a<0$) the
energy minimum in the bulk of the film is attained at either
$z_1=1$, $z_2=0$ or vice versa; we choose the first solution for
further discussions. Then the order parameter is proportional to
$e^{i(\chi-\varphi/2)}$ and an elementary vortex corresponds to
the rotation by $2\pi$ of the ``effective'' parameter's phase
$\chi -  \varphi/2 = \phi$, where $\phi$ is the azimuthal angle on
the plane.   Near the vortex core, however, one can construct
solutions which contain both $z_1$ and $z_2$ components. Now we
show that such a solution has a lower energy than that of a
standard  singular vortex (like those in He-II or strongly type-II
superconductors).

Indeed, one can employ a vortex trial solution
\begin{eqnarray}\label{BPanzats}
z_1=\frac{re^{i\phi}}{\sqrt{R^2+r^2}},
\qquad\quad z_2=\frac{R}{\sqrt{R^2+r^2}},
\end{eqnarray}
which satisfies the boundary conditions : only one component
$z_1=\frac{\Delta_+}{|\Delta|}=e^{i\phi}$ survives on large
distances. The solution~(\ref{BPanzats}) possesses a topological
charge ${\cal Q}=1$, and it is just the
skyrmion~[\onlinecite{Skyrm}, \onlinecite{BP}] for the $N=3$
non-linear sigma model functional (i. e. the first term in the
free energy (\ref{includesigma}))
\begin{equation}
E_{\rm Skyrm}=
\frac{\rho_s}{2}\int\frac{1}{4}(\partial_{\mu}\vec{n})^2d^2\vec{r}.
\label{N3}
\end{equation}
Note, however, that here we use a different gauge. The
solution~(\ref{BPanzats}) corresponds to choice of the phases
\begin{equation}
\chi = -\varphi/2 = \phi/2.
\label{chiphi}
\end{equation}
The parameter $R$ is an arbitrary number: at any $R$ the topological
charge of the skyrmion is ${\cal Q}=1$ and its energy
$$E_{\rm Skyrm} = \pi\rho_s.$$
However, the full gradient functional (\ref{includesigma})
contains the second term as well.  This term leads  to
logarithmically large energy
\begin{equation}
E_2 = \pi\rho_s \log\frac{\Lambda}{R} \, ,
\label{E2}
\end{equation}
where $\Lambda$ is the minimal of the system size and (a very long)
two-dimensional screening length $\lambda_{2D} = 2\lambda^2/d$.
Indeed, at $ r \geq R$ one finds ${\cal A}_\mu^2 = r^{-2}$, whereas
at $r \ll R$ the polar angle $\theta \to \pi$ and,
according to Eqs.~(\ref{Aa}, \ref{chiphi}) the ``vector potential''
${\cal A}_\mu$ is not singular anymore.
It is evident from Eq.~(\ref{E2}) that one should choose $R$ as large
as possible in order to minimize vortex energy.  The upper limit is
given by the anisotropy length $L_{\rm an}$ defined in Eq.~(\ref{Lan}).
Thus the minimal energy of our trial solution can be estimated as
\begin{equation}\label{continuous}
E_{cont} =
\pi\rho_s
  (\log{\Lambda/L_{\rm an}}+ C ) \, ,
\end{equation}
where $C \sim 1$. The energy of the continuous
vortex~(\ref{continuous}) is lower than the energy of a usual
singular vortex by a large amount
\begin{equation}
E_{sing} - E_{cont} = \frac{\pi}{2}\rho_s\log\frac{\beta_s}{|\beta_a|}.
\label{delta}
\end{equation}
We emphasize, that the solution (\ref{BPanzats}) does not provide
the energy minimum but is just a trial function; the correct
non-singular vortex solution should have even lower energy and
thus is more stable than the singular vortex. In the case of a
continuous vortex the term $C_0$ in Eq.~(\ref{indent}) is zero and
${\cal Q}=\Phi/\Phi_0$, whereas for a singular vortex $C_0 = \pm
1$ and ${\cal Q}=0$.

\subsection{Half-quantum vortices in the LOFF state}

The LOFF-like (or ``stripe'') state is realized at $\beta_{a}>0$
and its degeneracy manifold is  $S^1 \otimes S^1$. Indeed, the
energy minimum is realized when vector $\vec{n}$ is parameterized
as $\vec{n}=(\cos\varphi,\sin\varphi,0)$, i. e. $\theta =\pi/2$,
thus there are two phase variables, $\varphi$ and $\chi$, cf.
parameterization (\ref{zEuler}, \ref{nEuler}). A usual singular
vortex solution corresponds to
 $\varphi =$ const and  $\delta\chi = 2\pi$;
according to Eq.~(\ref{Aa}), in this case the ``vector potential''
${\cal A}_\mu = \partial_\mu\chi$, and it does not contain the
second phase $\varphi$. Then a natural question arises, if some
other vortex-like solutions are possible, due to the extended
(with respect to the usual $S^1$ ) degeneracy manifold. Indeed,
the same degeneracy manifold of the order parameter is realized in
some of the p-wave superconductive states, leading to the
existence of half-quantum
vortices~\cite{VolMin76,Volovik99,Ivanov00}. The reason for the
existence of such an object is evident from the representation
(\ref{zEuler}): a sign-change of the order parameter due to the
$\pi$-rotation of the phase $\chi$ along some closed loop in real
space can be compensated by the $\pm 2\pi$-rotation of the phase
$\varphi$ along (topologically) the same loop.

We are not aware of any explicit solution for a half-vortex in the
general case of an $S^1 \otimes S^1$ degeneracy manifold. However,
some progress can be made in the vicinity of the symmetric point
${\cal S}$, where $\beta_a \ll \beta_s$ and the problem simplifies
a bit due to the presence of the ``isotropic'' spatial scales,
$\xi(T) \ll L \ll L_{\rm an}(T)$, cf. Eq.~({\ref{Lan}).  Indeed,
on such length-scales the problem can be treated within the
gradient free energy, cf. Eq.~(\ref{4N}) or
Eq.~(\ref{includesigma}).  The solutions with half-quantum of
magnetic flux  obey boundary conditions $\theta(\infty) = \pi/2$
and $\delta_{\infty}\varphi = \pm 2\pi$, where via
$\delta_{\infty}$ we denoted the phase increment along the large
loop.  Explicit form of these solutions may be ($\tilde{z_i} =
\sqrt{2}z_i$)
\begin{eqnarray}
\tilde{z}_1 = \sqrt{1-\frac{R}{\sqrt{r^2+R^2}}}
e^{ i\gamma\phi}, \, \quad
\tilde{z}_2 = \sqrt{1+\frac{R}{\sqrt{r^2+R^2}}}  \nonumber  \\
{\rm and} \nonumber \\
\tilde{z}_1 = \sqrt{1+\frac{R}{\sqrt{r^2+R^2}}}, \, \quad
\tilde{z}_2 = \sqrt{1-\frac{R}{\sqrt{r^2+R^2}}}e^{i\gamma\phi}, \,
\label{z-half-vortex}
\end{eqnarray}
with an arbitrary parameter $R$. The variable $\gamma = \pm 1$ in
the exponent of Eq.~(\ref{z-half-vortex}) corresponds to the sign
of the vorticity (the magnetic flux), whereas the first and the
second lines in Eq.~(\ref{z-half-vortex}) correspond to the cases
of negative and  positive projections $n_3$ of the $\vec{n}$
vector in the center of the half-vortex. In terms of vector
$\vec{n}$ these four solutions (\ref{z-half-vortex}) are
represented as follows:
\begin{equation}\label{n-half-vortex}
\vec{n}_{1,2} = \frac{1}{\sqrt{r^2 + R^2}}
\left(\begin{array}{c} x\\
                      -\gamma y\\
- R
\end{array} \right),\, \, \,
\vec{n}_{3,4} = \frac{1}{\sqrt{r^2 + R^2}}
\left(\begin{array}{c} x\\
                      \gamma y\\
 R
\end{array} \right) \, .
\end{equation}
An elementary calculation of the topological charge and of the
magnetic flux associated with the vortex solution
Eq.~(\ref{n-half-vortex}) leads to
\begin{equation}
{\cal Q} =  \frac{\Phi}{\Phi_0} = \frac{\gamma}{2}.
\label{Q2}
\end{equation}
An additional binary variable which characterizes the half-vortex
is the sign of the component $n_3(r=0)$. Therefore, totally there
are 4 types of half-quantum vortices.

In the presence of a finite length $L_{\rm an}$ the
solutions~(\ref{z-half-vortex}) make sense as intermediate
asymptotics, as long as $R \ll L_{an}$, whereas at longer scales
the anisotropic term $\propto \beta_a$ modifies the solution
considerably (a numerical solution would be necessary to determine
the solution in this region). Nevertheless, we can make an energy
comparison between the singular vortex and a pair of half-quantum
vortices even without an explicit solution including the
anisotropic term. Consider two contributions to the free energy
functional~(\ref{includesigma}). The term with $\partial_\mu{\vec
n}$, evaluated for the solution (\ref{n-half-vortex}), gives (the
main contribution comes from large distances, $r \gg R$)
$$E_{\vec{n}} \approx \frac{\pi}{4}\rho_s \log\frac{\Lambda}{R},$$
whereas the term with ${\cal A}_\mu$ contributes with about the
same amount (since ${\cal A}_\mu$ is non-singular at small $r \leq
R$)
$$E_{\cal A} \approx
\frac{\pi}{4}\rho_s\log\frac{\Lambda}{R}.$$
Both above estimates
may contain subleading terms $\sim \rho_s$ which we do not
control. Totally, the energy of a half-vortex is
\begin{equation}
E_{1/2} = \frac{\pi}{2}\rho_s\left(\log\frac{\Lambda}{R}  +
C_{1/2}\right),
\label{halfenergy}
\end{equation}
where $C_{1/2} \sim 1$. The minimal energy of a half-vortex can be
estimated from Eq.~(\ref{halfenergy}) with the substitution $R\sim
L_{\rm an}$.  Thus we find that the energy of two half-vortices
coincides (up to the terms $\sim \rho_s$ which do not contain a
large logarithm) with the energy of a continuous single-quantum
vortex (\ref{continuous}), found in the previous Subsection, and
is certainly lower than the energy of the singular vortex by
approximately the same amount as in Eq.~(\ref{delta}). This means
that the half-quantum vortex is a fundamental topological defect
of the stripe (LOFF) state, at least in the region relatively
close to the symmetric point ${\cal S}$.

\section{Phase diagram}
\label{PhaseDiagram}

\subsection{Stationary conditions for the helical phase}
\label{Stationary}

Now we concentrate on the properties of the helical phase
significantly below $T_c(H)$, and determine the locations of the
phase transition lines ${\cal LT}$, ${\cal ST'}$ and ${\cal TO}$.
This calculation is possible since $|\Delta(\vec{r})|^2 = {\rm
const}$ in the helical state, and thus explicit analytic equations
determining $\Delta$ and the corresponding $Q$ can be written
without resorting to an expansion over small $\Delta$. Evaluation
of the thermodynamic potential in the helical state gives
\begin{eqnarray}
\label{SCenergy}
\Omega_{hel}(\Delta,H)=\frac{\Delta^2}{U} -
\qquad\qquad\qquad\qquad\qquad\qquad\qquad \\
\pi\nu(\epsilon_F) T\sum_{\omega,\lambda} \int_0^{2\pi}
\left(\sqrt{\tilde\omega^2+\Delta^2} -
|\tilde\omega|\right)\frac{d\varphi}{2\pi},\nonumber
\end{eqnarray}
where
\begin{eqnarray}\label{tildeomega}
\tilde\omega=\omega+iH_\lambda\sin\varphi,
\end{eqnarray}
and $H_\lambda$ is determined in Eq.~(\ref{Hlambda}).
The equations determining  $\Delta$ and $Q$ are derived from the
two stationary conditions
\begin{eqnarray}\label{stationarygap}
\frac{\partial\Omega_{hel}}{\partial \Delta}=\frac{2\Delta}{U} -
\pi\nu(\epsilon_F) T\sum_{\omega,\lambda}
\int_0^{2\pi}\frac{\Delta}{\sqrt{\tilde\omega^2+\Delta^2}}
\frac{d\varphi}{2\pi}=0\quad
\end{eqnarray}
and
\begin{eqnarray}\label{stationaryQ}
&&\frac{\partial \Omega_{hel}}{\partial Q}=
\frac{v_F}{2}\nu(\epsilon_F)T\sum_{\omega,\lambda}
f(H_\lambda,\omega)=0,
\end{eqnarray}
where we denoted
$f(H_\lambda,\omega)=-i\int_0^{2\pi}
\frac{\tilde\omega\sin\varphi}
{\sqrt{\tilde\omega^2+\Delta^2}}
\frac{d\varphi}{2}.$

Reducing the integrals over $\varphi$ in
Eqs.~(\ref{stationarygap}, \ref{stationaryQ}) to complete elliptic
integrals (see Appendix~\ref{Gap}) compacts the stationary
conditions to a two equations set
\begin{eqnarray}\label{secline}
\frac{1}{U}=2\nu(\epsilon_F)T\sum_{\omega>0,\lambda}\frac{{\bf K}
\left(k\right)}{\sqrt{\omega^2+(|H_\lambda|+\Delta)^2}},
\end{eqnarray}
\begin{eqnarray}
\sum_{\omega>0,\lambda}f(H_\lambda,\omega) = 0,
\label{2cond}
\end{eqnarray}
where the Jacobi modulus
\begin{eqnarray}\label{Jacobimodulus}
k=\frac{2\sqrt{\Delta |H_\lambda|}}{\sqrt{\omega^2+
(|H_\lambda|+\Delta)^2}},
\end{eqnarray}
and the function
\begin{eqnarray}\label{fHlambda}
&&f(H_\lambda,\omega)=\frac{1}{H_\lambda}
\Big((\omega^2+H_\lambda^2+\Delta^2)\frac{{\bf K}(k)}
{\sqrt{\omega^2+(|H_\lambda|+\Delta)^2}}-
\nonumber \\
&&-\sqrt{\omega^2+(|H_\lambda|+\Delta)^2}{\bf E}(k)\Big)
\end{eqnarray}
is defined through the Jacobi complete elliptic integrals of the
first and second kind. At $\Delta=0$ Eq.~(\ref{secline}) reduces
to Eq.~(\ref{T_c(H)}).

\subsection{Lifshitz phase transition line ${\cal{LT}}$}
\label{Lline}

The thermodynamic potential~(\ref{SCenergy}) in the helical state
is a function of $Q$ for any given pair $(H,T)$ by virtue of
Eqs.~(\ref{secline}, \ref{2cond}). At small $Q$
Eq.~(\ref{SCenergy})  can be expanded in powers of $Q$ as (terms
of the order of $\alpha/v_F$ have been neglected)
\begin{equation}
\Omega_{hel}(Q)=\Omega_{hel}(0)+aQ^2+bQ^4+cQ^6,
\label{Omegaexp}
\end{equation}
where
\begin{eqnarray}\label{afirst}
a\!=\!\frac{1}{2}\frac{\partial^2\Omega}{\partial Q^2}
=\left(\frac{v_F}{2}\right)^2\!\nu(\epsilon_F)T
\sum_{\omega,\lambda}\int_0^{2\pi}\!\frac{\Delta^2\sin^2{\varphi}}
{\left(\tilde\omega^2+\Delta^2\right)^{3/2}}
\frac{d\varphi}{2},\nonumber \\
\end{eqnarray}
\begin{eqnarray}\label{b}
24 b=\frac{d^4E}{dQ^4}=\frac{\partial^4\Omega}{\partial Q^4}-
3\frac{\left(\frac{\partial^3\Omega}{\partial\Delta\partial Q^2}
\right)^2}
{\frac{\partial^2\Omega}{\partial\Delta^2}}
\end{eqnarray}
and $c>0$. The condition $a=0$, $b>0$ determines the second-order
Lifshitz phase transition line ${\cal LT}$, which ends at the
critical point $\cal T$, where the coefficient $b=0$ changes sign,
see Fig.~\ref{phd}. The condition $a=0$ may be simplified after
reducing the integrals over $\varphi$ in Eq.~(\ref{afirst}):
\begin{eqnarray}\label{KKK}
&&\sum_{\omega>0}
\left( J+2\omega \frac{\partial}{\partial \omega}J +
\frac{\Delta^2-\omega^2}{\Delta}\frac{\partial}
{\partial \Delta}J\right)=0, \\ \nonumber
&&\mbox{where }\\ \nonumber
&&J=\frac{{\bf K}
\left(\sqrt{\frac{4\Delta H}{\omega^2+(H+\Delta)^2}}\right)}
{\sqrt{\omega^2+(H+\Delta)^2}}.
\end{eqnarray}
The line  ${\cal LO}$ of stability of the BCS state (with respect
to a formation of the helical wave) is determined by a
simultaneous numerical solution of Eqs.~(\ref{secline},
\ref{2cond}),  taken at $Q=0$, and Eq.~(\ref{KKK}). This line is
indeed a line of a second-order transition as long as the
coefficient $b > 0$. Using  Eqs.~(\ref{SCenergy}, \ref{2cond}), we
compute the coordinates of the point ${\cal T}\in {\cal{LO}}$
where $b=0$ (cf. Eq.~(\ref{b}) for $b$) as $(H,T)=(1.547,
0.455)T_{c0}$. At lower temperatures $b < 0$ and a first-order
transition out of the homogeneous state takes place. Therefore
${\cal TO}$ is a boundary of a domain of the BCS state local
stability. The actual first-order transition line ${\cal TO'}$
between the BCS and some inhomogeneous state consisting of many
spatial harmonics lays at lower values of the magnetic field
($H_{O'}<H_O=1.76T_{c0}$).

The homogeneous superconductor which exists
on the left of the Lifshitz line
is ``gapless'' for high enough temperatures.
The spectrum of the particles is
$E_{\bf p}=\sqrt{\xi^2+\Delta^2}-\lambda H\sin{\varphi}_{\bf p}$,
therefore the minimum bound energy of the Cooper pairs
(the energy gap) turns to zero
at $H\geq \Delta$.
The line of transition into the gapless superconductivity
$H=\Delta$ is marked in Fig.~\ref{phd} with a dashed line.

\subsection{Phase transition line ${\cal ST'}$}
\label{Sline}

The second-order phase transition line $\cal {ST'}$ bounds the region
 of the helical phase from the high-$H$ side. Its position
can be determined via the stability condition with respect to the
additional  modulations of the order parameter, of the form
$\delta \Delta(\vec{r})= \delta v_{-q}\exp(-iqy) + \delta
v_{q+2Q}\exp(i(q+2Q)y)$:
\begin{eqnarray}
\delta\Omega_{\delta v} = \vec{v}^+\hat{\cal A}(q) \vec{v},
\end{eqnarray}
where $\vec{v}=(\delta v_{-q},\delta v^*_{q+2Q})$ and
\begin{eqnarray}
\hat{\cal A}(q)\!=\!\frac{\hat{1}}{U}-
\!\!\sum_{\omega>0,{\bf p},\lambda}\!
\left( \begin{array}{cc} G_{\lambda p_-}
G_{\lambda -p_+} &  F_{\lambda p_-}
F_{\lambda -p_+} \\  F^*_{\lambda p_-}
F^*_{\lambda -p_+} &  G_{\lambda p_++Q}
G_{\lambda -p_-+Q} \end{array}\right),
\nonumber \\
\end{eqnarray}
whith $p_{\pm}=p\pm q/2$. The Green's functions entering the
matrix $\hat{{\cal A}}$ are
\begin{eqnarray}
&&G_\lambda \left(\omega,{\bf p}-\frac{{\bf q}}{2}\right)=
-\frac{i\tilde\omega+\xi-{\cal Q}_{\bf p}\quad\qquad}
{\tilde\omega^2+(\xi-{\cal Q}_{\bf p})^2+\Delta^2},
\nonumber \\
&&F_\lambda \left(\omega,{\bf p}-\frac{{\bf q}}{2}\right)=
\frac{\lambda e^{-i\varphi_{\bf p}}\Delta}
{\tilde\omega^2+(\xi-{\cal Q}_{\bf p})^2+\Delta^2}=
\nonumber \\
&&=-F_\lambda \left(-\omega,-{\bf p}+\frac{{\bf q}}{2}+{\bf Q}\right),
\qquad\qquad\qquad\qquad\qquad
\end{eqnarray}
where $\tilde\omega$ is given by~(\ref{tildeomega}), and for
brevity we introduced a notation ${\cal Q}_{\bf
p}=(q+Q)\sin{\varphi_{\bf p}}/2$.

The matrix $\hat{{\cal A}}$ has two eigenvalues
$\epsilon_1(q)<\epsilon_2(q)$,
\begin{eqnarray}\label{eigen}
&&\epsilon_{2,1}(q)=\nonumber \\
&&\!=\!\!\left(\frac{1}{U}-\frac{g_{-q}+g_{q+2Q}}{2}\right)\pm
\sqrt{\left(\frac{g_{-q}-g_{q+2Q}}{2}\right)^2+\left|f_{-q}\right|^2},
\nonumber \\
&&\mbox{ where} \nonumber \\
&&\frac{1}{U}-\frac{g_{-q}+g_{q+2Q}}{2}= \nonumber \\
&&=\sum_{\omega>0,\lambda}\int_0^{2\pi}\frac{d\varphi}{4}
\frac{1}{\sqrt{\tilde\omega^2+\Delta^2}}
\left (1-\frac{\tilde\omega^2-
{\cal Q}_{\bf p}^2}
{\tilde\omega^2+
{\cal Q}_{\bf p}^2+\Delta^2}
\right ),\nonumber \\
&&\frac{g_{-q}-g_{q+2Q}}{2}\!=
\!\!\sum_{\omega>0,\lambda}\int_0^{2\pi}\frac{d\varphi}{4}
\frac{1}{\sqrt{\tilde\omega^2+\Delta^2}}\,
\frac{2i\tilde\omega{\cal Q}_{\bf p}}
{\tilde\omega^2+{\cal Q}_{\bf p}^2+\Delta^2},
\nonumber \\
&&f_{-q}=\sum_{\omega>0,\lambda}
\int_0^{2\pi}\frac{d\varphi}{4}\frac{1}
{\sqrt{\tilde\omega^2+\Delta^2}}\,
\frac{\Delta^2}{\tilde\omega^2+{\cal Q}_{\bf p}^2+\Delta^2}.
\nonumber \\
\end{eqnarray}
Here $g_{-q}=\sum_{\omega>0,{\bf p},\lambda}
G_{\lambda,{\bf p}-{\bf q}/2}G_{\lambda,-{\bf p}-{\bf q}/2}$,
$g_{q+2Q}=\sum_{\omega>0,{\bf p},\lambda}
G_{\lambda,{\bf p}+{\bf q}/2+{\bf Q}}
G_{\lambda,-{\bf p}+{\bf q}/2+{\bf Q}}$
and $f_{-q}=\sum_{\omega>0,{\bf p},\lambda}
F_{\lambda,{\bf p}-{\bf q}/2}F_{\lambda,-{\bf p}-{\bf q}/2}$.
The integrals (\ref{eigen}) can be expressed through the
elliptic integrals of the first and the second kind,
as shown in Appendix~\ref{ST'}.

The helical state metastability line ${\cal ST'}$ is defined as a
set of points where one mode $\delta v$ becomes energetically
favorable: ${\rm min}_q\epsilon_1(q)=0$. Four equations are to be
solved simultaneously in order to find the position of this line:
two gap equations~(\ref{secline}, \ref{2cond}), that determine the
equilibrium $\Delta$ and $Q$, together with the two equations
$\partial_q\epsilon_1(q)=0$ and $\epsilon_1(q)=0$, where
$\epsilon_1(q)$ is determined in~(\ref{eigen}). We call the new
phase on the right of the ${\cal{ST'}}$ transition line a
``three-exponential'' state.

Note, that the ${\cal ST'}$ line is an actual phase transition
line out of the helical state {\it if} this transition is of the
second order. Another possibility might be that  a first-order
transition occurs, which transforms the helical state into
parity-even LOFF-type state, and occurs at slightly lower values
of $H$ at each $T$. Below we show that near $T_c(H)$ the phase
transition is indeed of the second order.  In order to demonstrate
it, we evaluate terms of the eighth order in $|\Delta_Q|$ in the
Ginzburg-Landau functional. Parameterizing the order parameter
spinor near the symmetric point as
\begin{equation}\label{LLeq1}
\left(\begin{array}{c} \Delta_+\\ \Delta_-\end{array} \right)=
\Delta \,e^{i\chi}
 \left(\begin{array}{c} e^{-i\varphi/2}\cos{\theta/2}\\
e^{i\varphi/2}\sin{\theta/2}\end{array} \right),
\end{equation}
the anisotropy part in the Ginzburg-Landau functional
reads
\begin{equation}\label{Fanis}
\Omega_{sn}^{anis}(\kappa)=\beta_a \Delta^4\cos^2{\theta}+
\kappa\Delta^8\cos^4{\theta}
\end{equation}
(note that the term $\propto |\Delta|^6$ is not allowed by
symmetry). The sign of the coefficient $\kappa$ in front of the
term $\cos^4{\theta}$ determines the type of the transition near
the $T_c(H)$ line. We demonstrate, by rather tedious  calculations
described in Appendix~\ref{8}, that $\kappa>0$, which proves that
the transition is of the second order near the $T_c(H)$ line.

Now we come to somewhat surprising situation. Indeed, according to
the analysis of superconductive instability at the $T_c(H)$ line,
on the right of the symmetric point $\cal{S}$ the stripe (
LOFF-like) state is formed, which contains two harmonics with $ Q
= \pm |Q|$. Such a phase preserves spatial inversion (in the
plane) and time-reversal symmetry. Below the  $T_c(H)$ line
additional harmonics develop in such a phase, but they come in
pairs $\pm 3Q,\pm 5Q,...$, and still preserve spatial and
time-reversal symmetry. On the other hand, the second-order phase
transition line $\cal{ST'}$ separates the helical state
(parity-odd) and a ``three-exponential'' state which has broken
parity and time-reversal symmetry as well. It means that there
must exist one more phase transition line, between the
``three-exponential'' phase and the parity-even stripe phase. This
transition line should start at point $\cal{S}$ but will lie
slightly to the right from the line $\cal{ST'}$, i. e. at higher
values of the field.

The fact that the points $\cal T$ and ${\cal T}'$ are different
but close to each other is in favor of existence of a critical
point ${\cal K}$ on the line ${\cal ST'}$ (similar to the point
${\cal T}$ on the line ${\cal LO}$) below which the phase
transition from the helical to the LOFF-type state becomes a
weakly first order transition.

\section{Current and electromagnetic response in the helical state}
\label{Current}

\subsection{The absence of the equilibrium current in the
helical ground-state}
\label{ZeroCurrent}

The oscillating space-dependence of the order parameter like given
by Eq.~(\ref{DeltaHarmExp}) may lead to  the hypothesis of a
nonzero background electric current present in such a state. We
will show here by general arguments that the electric current is
in fact absent in the helical state: the condition of its
vanishing is equivalent to the minimum of the free energy with
respect to the variation of the structure's wavevector $Q$.

The superconducting current can be written in the following form:
\begin{eqnarray}\label{currentoperator}
&&\vec{j}_s=\frac{e}{2}T\sum_{\omega,{\bf p},\lambda}
\Big(\frac{\partial\epsilon_{\lambda,{\bf p}+{\bf Q}/2}}
{\partial{\bf p}}
G_{\lambda\lambda}(\omega,{\bf p}+{\bf Q}/2)-\nonumber \\
&&-\frac{\partial\epsilon_{\lambda,-{\bf p}+{\bf Q}/2}}
{\partial{\bf p}}
G_{\lambda\lambda}(-\omega,-{\bf p}+{\bf Q}/2)\Big),
\end{eqnarray}
where
\begin{eqnarray}\label{generalGreenfunction}
G_{\lambda\lambda}(\omega,{\bf p}+{\bf Q}/2)=
-\frac{i\omega+\epsilon_{\lambda,-{\bf p}+{\bf Q}/2}}
{\tilde{\omega}^2+\xi_\lambda^2+\Delta^2}
\end{eqnarray}
is the electron Green's function in the helical state;
$\epsilon_{\lambda,{\bf p}}$ is
the spectrum of the free electron. For brevity we used notations
$$\tilde{\omega}=\omega+i\frac{\epsilon_{\lambda,{\bf p}+{\bf Q}/2}-
\epsilon_{\lambda,-{\bf p}+{\bf Q}/2}}{2},$$
$$\xi_\lambda=\frac{\epsilon_{\lambda,{\bf p}+{\bf Q}/2}+
\epsilon_{\lambda,-{\bf p}+{\bf Q}/2}}{2}.$$
The thermodynamic potential in the helical state
\begin{eqnarray}
\Omega_{hel} =-\frac{T}{2}\sum_{\omega,{\bf p},\lambda}
\log{}\left[\tilde{\omega}^2+\xi_\lambda^2+\Delta^2\right]
\end{eqnarray}
is evaluated
explicitly
for an arbitrary spectrum $\epsilon_{\lambda,{\bf p}}$
of the electron, due to $|\Delta(\vec{r})|=$const.

The derivative is
\begin{eqnarray}\label{generalstationarycondition}
&&\frac{\partial\Omega_{hel}}{\partial \vec{Q}}\!=\!\\
&&-\frac{T}{4}\!\sum_{\omega,{\bf p},\lambda}
\frac{\frac{\partial\epsilon_{\lambda,{\bf p}+{\bf Q}/2}}
{\partial\bf p}
(i\tilde{\omega}+\xi_\lambda)-
\frac{\partial\epsilon_{\lambda,{-\bf p}+{\bf Q}/2}}
{\partial\bf p}
(-i\tilde{\omega}+\xi_\lambda)}
{\tilde{\omega}^2+\xi_\lambda^2+\Delta^2}.\nonumber
\end{eqnarray}
Comparing this stationary condition~(\ref{generalstationarycondition})
with the expression for the current~(\ref{currentoperator}), we notice
\begin{equation}\label{zerocurrent}
\vec{j}_s=\frac{2e}{\hbar}\frac{\partial
\Omega_{hel}}{\partial\vec{Q}}.
\end{equation}
Thus, a direct calculation of the superconducting current
$\vec{j}_s$ shows that in any order of $\alpha/v_F$ the current in
equilibrium is zero. This result apparently contradicts the
statement made by Yip~\cite{Yip}, who studied basically the same
model as the present one, and found a non-zero supercurrent in the
presence of a parallel magnetic field. The resolution of this
paradox will be presented in Sec.~\ref{WeakHelical} below.

\subsection{Electromagnetic response in the helical state}
\label{EMresponse}

We calculated the electromagnetic response function
$\delta j_\alpha/\delta A_\beta = -\frac{e^2}{m c}n_s^{\alpha\beta}$
for the helical state using the standard diagram methods:
\begin{eqnarray}
&&j_x=\frac{e^2}{c}T\sum_{\omega,{\bf p},\lambda}
\left(G^2+F^2\right)\cos^2\varphi
\left(\frac{p}{m}-\lambda \alpha\right)^2 A_x, \nonumber \\
&&j_y=\frac{e^2}{c}T\sum_{\omega,{\bf p},\lambda}
\left(G^2+F^2\right)\sin^2\varphi
\left(\frac{p}{m}-\lambda \alpha\right)^2 A_y,\nonumber \\
\end{eqnarray}
where $G=G(\omega,{\bf p}+{\bf Q}/2)$ and
$F=F(\omega,{\bf p}+{\bf Q}/2)$ are correspondingly the normal and
the anomalous Green's functions in the helical state.
We found that
\begin{eqnarray}\label{SCdensity}
n_s^{yy} = 4\frac{m}{\hbar}\frac{\partial^2\Omega}{\partial Q^2},
\end{eqnarray}
i. e. proportional to the parameter $a$ from the expansion of the
thermodynamic potential~(\ref{Omegaexp}). Thus on the Lifshitz
line $\cal{L}\cal{T}$ there is no linear supercurrent in the
direction perpendicular to the magnetic field. The component
$n_s^{xx}$ does not vanish anywhere in the helical state region
and is of the order of $n_s$ of the BCS state:
\begin{equation}
n_s^{xx} = \frac{4m}{\hbar^2}
\left(\frac{v_F^2}{2}
{\Delta\frac{\partial}{\partial \Delta}
\left(\frac{1}{\Delta}\frac{\partial\Omega_{hel}}
{\partial \Delta} \right)-\frac{\partial^2 \Omega_{hel}}
{\partial Q^2}}\right).
\end{equation}
This  anisotropic behavior of the superfluid density tensor is in
contrast with the one found in the  classical LOFF problem, where
$n_s$ was shown to vanish in the whole helical state; the
difference is probably due to the fact that in our problem the
direction of ${\bf Q}$ is fixed by the applied field ${\bf h}$,
while in the case of a ferromagnetic superconductor it is
arbitrary. The obtained behavior of the $n_s^{\alpha\beta}$ tensor
indicates a strongly anisotropic electromagnetic response of the
surface superconductor near the Lifshitz line $\cal{L}\cal{T}$.

\section{``Weakly helical'' phase at low magnetic fields}
\label{WeakHelical}
\subsection{Transformation of the uniform BCS state into
long-wavelength helical state} \label{Qsmall}

The thermodynamic potential of the helical state in the form of
Eq.~(\ref{SCenergy}) was obtained while neglecting the term
$\alpha Q$ small in comparison with the term $v_F Q$ in the energy
of the electron. Within this approximation the thermodynamic
potential was symmetric under the change $Q\rightarrow -Q$ and
hence the expansion~(\ref{Omegaexp}) contained only even powers of
$Q$. The account for the first-order term in $\alpha/v_F$ in
expressions of the type of Eq.~(\ref{SCenergy}) leads (see below)
to an appearance of the term  $\eta Q$  in the thermodynamic
potential (\ref{Omegaexp}). This term results in the
transformation of the homogeneous BCS state (situated on the left
of the $\cal {LT}$ line)  into a weakly helical phase with a small
wave-vector $|Q_{hel}|=\frac{2\alpha H}{v_F^2}$, first found
in~[\onlinecite{Agterberg}] for small values of $H$.
Mathematically, the terms of the order of $\alpha/v_F$ are
considered via taking into account the difference in the density
of states in Eq.~(\ref{SCenergy}):
$\nu(\epsilon_F)\rightarrow\nu_{\lambda}(\epsilon_F)$. Then the
second stationary condition~(\ref{2cond}) is changed noticeably:
\begin{eqnarray}
\frac{\partial \Omega_{hel}}{\partial Q}=
\nu(\epsilon_F)T\sum_{\omega,\lambda}\left(\frac{v_F}{2}+
\lambda\frac{\alpha}{2}\right)
f(H_\lambda,\omega)=0,
\end{eqnarray}
where $ f(H_\lambda,\omega)$ is defined in Eq.(\ref{fHlambda}).
Now the $Q=0$ solution never (at any $H \neq 0$)
provides a minimum for the superconducting energy:
\begin{eqnarray}\label{OmegaQ}
\eta=\left.\frac{\partial \Omega_{hel}}{\partial Q}\right |_{Q=0}=
\alpha\nu(\epsilon_F)T\sum_{\omega}f(H,\omega)\neq 0,
\end{eqnarray}
and the expansion of the thermodynamic potential in powers of $Q$
contains the linear term:
\begin{equation}\label{expansion_with_eta}
\Omega_{hel}(Q)=\Omega_{hel}(0)+\eta Q+\tilde{a}Q^2+...\quad,
\label{Omegaexplinear}
\end{equation}
which clearly indicates
that the equilibrium wave vector modulating the order parameter
on the left of the $\cal{LT}$ line
is always non-zero, although small as $\alpha/v_F \ll 1$:
\begin{eqnarray}\label{Qtrue}
&&Q_{hel}=-\frac{\eta}{2\tilde{a}}
=-\frac{2\alpha H}{v_F^2}\times\\
&&\times\frac{\sum_{\omega}
\left[\frac{(\omega^2+H^2+\Delta^2)}{{\cal{V}}}{\bf K}(k)-
{\cal{V}}{\bf E}(k)\right]}{\sum_{\omega}
\left[-\frac{(\omega^2+\Delta^2)}{{\cal{V}}}{\bf K}(k)+
\frac{(\omega^2+\Delta^2)^2+H^2(\omega^2-\Delta^2)}
{{\cal{V}}\left(\omega^2+(H-\Delta)^2\right)}{\bf E}(k)\right]},
\nonumber
\end{eqnarray}
where $k$ is the Jacobi modulus (\ref{Jacobimodulus}) and
${\cal{V}}=\sqrt{\omega^2+(H+\Delta)^2}$. In the limit
$H\rightarrow 0$ Eq.~(\ref{Qtrue}) can be expanded, resulting in
the dependence linear in $H$:
\begin{equation}\label{Qhel}
Q_{hel}=-2\alpha H/v_F^2.
\end{equation}
The formula~(\ref{Qhel}) is applicable for small enough magnetic
fields; in particular, for $ T \approx 0.5 T_{c0}$ the field range
is limited by $H \leq 0.5T_{c0}$, cf. Fig.~\ref{Qeps}. At higher
fields the dependence of the pairing wave vector upon the magnetic
field becomes non-linear but still  can be approximated by
Eq.~(\ref{Qtrue}), until   $H \leq 1.3 T_{c0}$, cf.
Fig.~\ref{Qeps}. Within the above  range of magnetic fields the
expansion (\ref{Omegaexplinear}) of the thermodynamic potential is
still applicable.

When the magnetic field is further increased, the system
approaches the transition into a short-wavelength helical state
discussed in Secs.V and VI above. The major role of the small term
$\eta Q$ is to broaden the $\cal {LT}$ line of the second-order
transition from the uniform to the helical state into a narrow
crossover region. Now the  dependence of the pairing wave vector
upon the magnetic field should be obtained numerically by solving
the system of the two self-consistency
equations~(\ref{stationarygap})~and~(\ref{2cond}), while keeping
also the term $\propto \alpha/v_F$ via the substitution of
$\nu(\epsilon_F)$ by $\nu(\epsilon_F)(1+ \lambda \alpha/v_F)$. The
corresponding results for $Q(H)$ are shown by full thin lines in
Fig.~\ref{Qeps} for two values of $\alpha/v_F$.

\begin{figure}
\includegraphics[angle=0,width=3.4in]{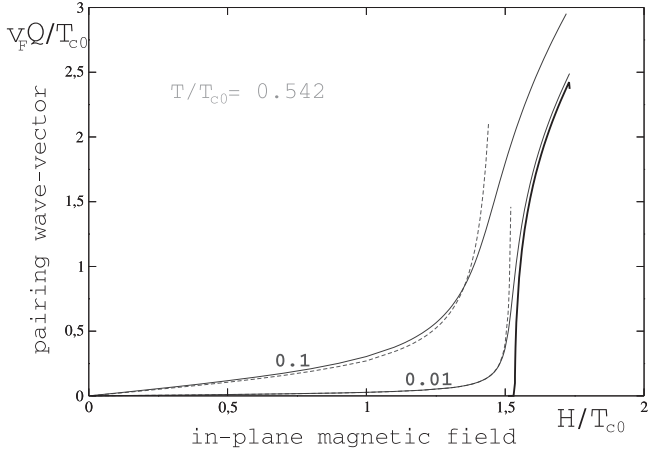}
\caption{\label{Qeps}
The dependence of the pairing wave vector $Q$ upon
the magnetic field for the two values:
$\alpha/v_F=0.1, 0.01$ (and a particular temperature
$T=0.54 T_{c0}$).
The curves shown by bold and thin lines correspond to
neglecting and taking into account
in the SC energy of the helical state
the small terms $\alpha/v_F$. The  dashed lines
correspond to the approximation~(\ref{Qtrue}). }
\end{figure}

\subsection{Cancellation of the ground-state current
and a spin-orbital analog of Little-Parks oscillations}
\label{LittleParks}

The gradient of the phase of the condensate wave function
determines the density of the superconducting current
\begin{equation}
{\bf j}_s^{(1)}=\frac{e \hbar}{2m}n_s\vec{Q}_{hel},
\end{equation}
where $n_s$ is the density of the number of superconducting
electrons, $e=-|e|$ is the charge of the electron, and $m$ is the
electron true mass. The expansion of Eq.~(\ref{SCdensity}) for
weak magnetic fields gives $n_s^{yy}=\frac{m
v_F^2}{\hbar}\nu(\epsilon_F)(1-Y(T,\Delta))$. In this limit the
density of the current in $y$-direction, induced by the
superconducting phase gradient, reads
\begin{equation}\label{j1}
j_y^{(1)}=-e\nu(\epsilon_F)\alpha (1-Y(T,\Delta))H.
\end{equation}
\begin{figure}
\includegraphics[angle=0,width=1.5in]{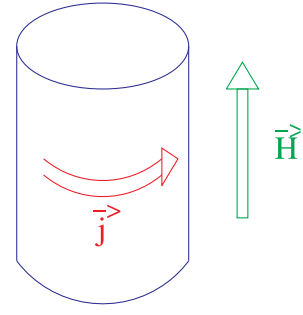}
\caption{\label{cylindr} A superconducting film wrapped
in a cylinder (cyclic boundary conditions).
This geometrical configuration enables
a current to flow in equilibrium in the ground state
of the helical phase when a parallel to the axis
magnetic field is applied.}
\end{figure}
\begin{figure}
\vskip0.1in
\includegraphics[angle=0,width=3.3in]{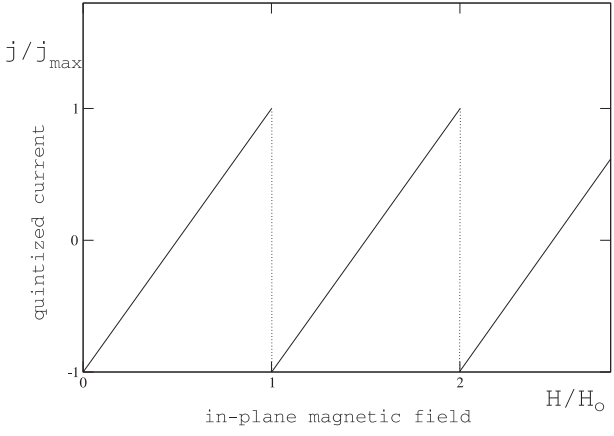}
\caption{\label{quantcurrent}
The sawtooth
 dependence of the superconducting current $j$,
flowing around a cylinder, upon the magnetic field $H$. The
amplitude is $j_{max}=\frac{e \hbar}{2m}n_s/R$, where $R$ is the
cylinder radius. The period is $H_{o}=\hbar v_F^2/(2\alpha R)$.
The strictly linear dependence, shown on the figure, takes place
in the case $H_o\ll T_{c0}$, $T=0$ only.}
\end{figure}
The supercurrent given by Eq.~(\ref{j1}) is not the only
contribution to be considered.   In fact, Yip~\cite{Yip} have
considered the same model and found that a weak current
proportional and perpendicular to the magnetic field flows in the
uniform BCS state. Indeed, a calculation of the current in the BCS
state gives a non-zero value (coinciding with that
in~[\onlinecite{Yip}])
\begin{equation}\label{j2}
j_y^{(2)}=T\sum_{\omega,{\bf p},\lambda}G_{\lambda}(\omega,{\bf p})
\hat{j}^{(chir)}_y=e\nu(\epsilon_F)\alpha (1-Y(T,\Delta))H.
\end{equation}
This second contribution leads to the presence (due to the Rashba
term in the Hamiltonian) of the anomalous contribution to the
electric current.

In the true ground-state with the weak helical modulations, both
contributions to the current, Eqs.~(\ref{j1}, \ref{j2}) sum up to
produce a perfect zero: $j_y^{(1)} + j_y^{(2)} = 0$, as they
should do according to the general proof given in
Sec.~\ref{ZeroCurrent} above. Thus we see  the resolution of the
controversy with the result~[\onlinecite{Yip}]: a uniform BCS
state considered by Yip does produce a supercurrent under the
action of a parallel magnetic field, but this state  {\it is not
the ground-state}.  Instead, the ground-state is realized as a
weakly helical state with a zero current. In fact, when a parallel
magnetic field is applied, the current does not flow, but a phase
difference $\Delta \chi = L Q_{hel}$ is induced at the edges of
the superconducting film in transverse to the field direction.

However, interesting ``traces'' of the spin-orbit-induced
supercurrent could possibly be seen, if the superconductor is
wrapped in a cylinder and a magnetic field is applied along the
axis. Then a current may flow around the cylinder (see
Fig.~\ref{cylindr}), due to the quantization condition $\delta
\chi=2\pi n,$ $n=0,\pm 1,\pm 2,...$, which does not allow an
arbitrary phase shift along the closed loop which encircles the
cylinder. The total current will be given by a sum of the
``spin-orbit current'' ${\bf j}_s^{(2)}$ and the current due to
the gradient of the quantized phase:
$$j_{quant}=\frac{e \hbar}{2m}n_s\left(-Q_{hel}+
\frac{2\pi n}{2\pi R}\right),$$
where $R$ is the cylinder radius and $Q_{hel}=-2\alpha H/v_F^2$.
The integer number $n$  is determined as a function of the
magnetic field in  a way to minimize the current: $n=$ integer
part of$[H/H_{o}]$, where $H_{o}=\frac{v_F^2}{2\alpha R}$. The
current will vanish at the field values $H=n H_{o}$, $n=0,\pm
1,\pm 2,...$ only. The maximal value of the current is equal (at
$T=0$) to $j_{max}=\frac{e \hbar}{2m}n_s/R$. The dependence of the
current on the magnetic field is of the sawtooth form, as shown in
Fig.~\ref{quantcurrent}, in the ideal case of zero temperature and
absence of impurities. The predicted  oscillations of the current
are, on first sight, similar to the well-known oscillations in a
superconducting cylinder (with the period $B_{LP} = \hbar c/eR^2$
in terms of a real magnetic induction), which goes back to the
early stages of the superconductivity studies.~\cite{LittleParks}
However, the corresponding oscillation periods differ by orders of
magnitude: $H_o/H_{LP} \sim (v_F/\alpha) \cdot (k_FR) \gg 1$. We
note also, that in a real experiment it should not be necessary
the superconducting film to be wrapped in a cylinder; it would be
sufficient to fabricate a heterostructure where a thin film with a
SO coupling would serve as a weak link introduced into a
SQUID-type loop, which would allow to control the superconducting
phase difference between the film edges.

\section{Phase diagram in the presence of non-magnetic impurities}
\label{Impurities}

In this Section we study the effects of potential impurities upon
the superconductive instability in the presence of the Rashba term
and the parallel magnetic field. We consider a standard model of
short-range weak impurities with a potential $u({\bf r}) =
u\delta({\bf r})$ and density $n_{imp}$; they are characterized by
an elastic scattering time $\tau$, where $\tau^{-1}=2\pi
n_{imp}u^2\nu(\epsilon_F)$. It is assumed that the impurity
scattering rate  is weak with respect to the Rashba splitting,
$1/\tau \ll \alpha p_F$, whereas it can be both weak or strong in
comparison with $T_c$.

The interaction between the electrons and the impurity is
described by the following Hamiltonian written in the chiral
representation:
\begin{eqnarray}\label{Himp}
\hat{H}_{int} =\frac{1}{V}\sum_i \sum_{{\bf p},{\bf p}'} u
e^{-i({\bf p}-{\bf p}'){\bf R}_i} a^+_{ {\bf p} \lambda}
M_{\lambda\mu} ({\bf p},{\bf p}')
a_{ {\bf p}'\mu},\nonumber \\
\end{eqnarray}
where the matrix
\begin{equation}\label{MM}
M_{\lambda\mu}({\bf p},{\bf p}')=\frac{1}{2}\left(1+\lambda\mu
e^{i(\varphi_p-\varphi_{p'})}\right)
\end{equation}
appears due to the transformation from the spin to chiral basis.
The semiclassical electron Green's function of the chiral metal in
the presence of non-magnetic impurities is
\begin{equation}
G_{\lambda}(\omega,{\bf p})=
\frac{1}{i\omega-\xi_\lambda({\bf p})-\lambda
H\sin{\varphi_{\bf p}}+\frac{i}{2\tau}{\rm sgn}{\omega}},
\label{Green_tau}
\end{equation}
where $\xi_\lambda({\bf p})$ is determined
in~(\ref{quasiparticle dispersion}).

The electron-electron vertex in the Cooper channel is then given
by non-crossing diagrams shown in Fig.~\ref{impureCooperloop},
where $\rm G$ denotes the Green's function~(\ref{Green_tau}). Note
that the chiralities of both electrons which belong to the same
``Cooper block'' coincide (otherwise, a diagram would be smaller
by a factor $(\alpha p_F\tau)^{-1} \ll 1$).

\begin{figure}
\includegraphics[angle=0,width=2.3in]{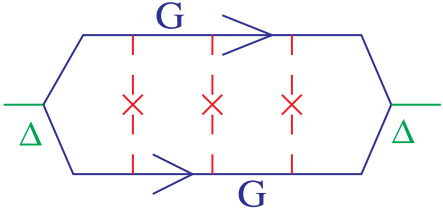}
\caption{\label{impureCooperloop}
The electron-electron vertex in the Cooper channel
in the presence of non-magnetic impurities.}
\end{figure}

\begin{figure*}
\includegraphics[angle=0,width=5in]{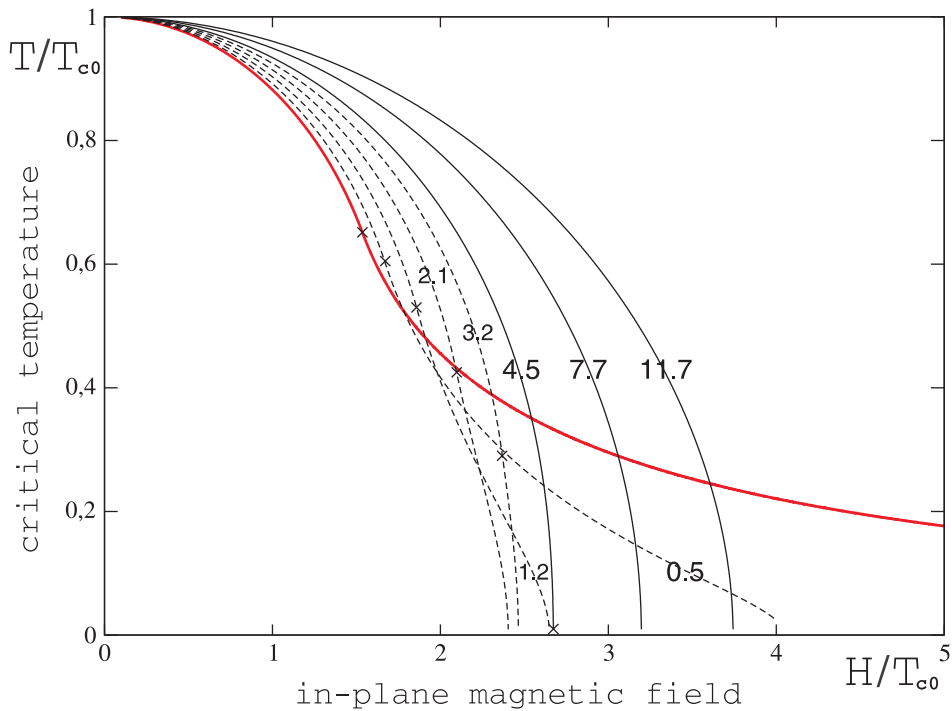}
\caption{\label{tau} Phase transition lines for different
strengths of impurity scattering: $1/2\tau T_{c0}= 0.0, 0.5, 1.2,
2.1, 3.2, 4.5, 6.0, 7.7, 9.6, 11.7.$ Lifshitz points are shown by
crosses.}
\end{figure*}

An analytical expression for the impurity line
is (cf. Eq.~(\ref{MM}) for the matrix
elements)
\begin{eqnarray}\label{amplitude}
V_{\lambda\mu}(\varphi_{\bf p},\varphi_{\bf p'})=
\frac{1}{8\pi\tau\nu(\epsilon_F)}
(1+\lambda\mu e^{i\varphi_{\bf p}-i\varphi_{\bf p'}})^2 =
\qquad\\ \nonumber
\frac{1}{4\pi\tau\nu(\epsilon_F)}  e^{i\varphi_{\bf p}-
i\varphi_{{\bf p}'}} (\lambda\mu+\sin{\varphi_{\bf p}}
\sin{\varphi_{{\bf p}'}}+
\cos{\varphi_{\bf p}}\cos{\varphi_{{\bf p}'}}),
\end{eqnarray}
where $\lambda$ and $\mu$ are the chiralities of the left and the
right blocks around the impurity line. In fact, the last term in
the r.h.s. of Eq.~(\ref{amplitude}) can be safely omitted since
its contribution vanishes after integration over the momenta in
the product of $V_{\lambda\mu}$ and the Cooper block,
Eq.~(\ref{Cooper_block}) below. Integrating over $\xi_\lambda$ a
block of two Green's functions (a ``Cooper block'') in the Cooper
channel gives
\begin{eqnarray}
&&C_{\lambda}(\omega,\sin{\varphi_{\bf p}})=\nonumber\\
&&=\nu_{\lambda}(\epsilon_F) \int_{-\infty}^{\infty} d\xi\
G_{\lambda}\left(\omega,{\bf p}+\frac{\bf q}{2}\right)
G_{\lambda}\left(-\omega,-{\bf p}
+\frac{\bf q}{2}\right)=\nonumber\\
&&=\frac{i\pi\nu_{\lambda}(\epsilon_{F})} {i\bar\omega
-H_{\lambda}\sin{\varphi_{\bf p}}},
\label{Cooper_block}
\end{eqnarray}
where $\bar\omega=\omega+{\rm sgn}\omega/2\tau$, $H_{\lambda}$ is
determined in~(\ref{Hlambda}) and
$\nu_{\lambda}(\epsilon_F)=\nu(\epsilon_F)(1+\lambda \alpha/v_F)$.

The superconductive transition temperature is determined by the usual
condition $U(0)^{-1} = {\cal C}$,
where ${\cal C}$ is the sum of all
ladder diagrams with $n=0, 1, 2,...$ impurity lines
(see Fig.~\ref{impureCooperloop}):
\begin{eqnarray}\label{C}
{\cal{C}}=
T\!\sum_{\omega>0}\!\sum_{n=0}^{\infty}\!\sum_{\lambda_n}\!
\int_0^{2\pi}\!
\frac{d\varphi_{{\bf p}_n}}
{2\pi} L^{n}_{\lambda_n}\!(\omega,\varphi_{{\bf p}_{n}}\!)
\times\nonumber\\
\times C_{\lambda_n}\!(\omega,\sin{\varphi_{{\bf
p}_n}}\!)\lambda_n e^{i\varphi_{{\bf p}_n}},
\end{eqnarray}
where $L^n_{\lambda_n}(\omega,\varphi_{{\bf p}_{n}})$ is the
expression for a ladder diagram which includes a left vertex $
L^0_{\lambda_0}(\omega,\varphi_{{\bf p}_{0}}) \equiv
\lambda_0e^{-i\varphi_{{\bf p}_0}}$, $n$ ``Cooper blocks'' and $n$
impurity lines; the factor $\lambda_n e^{i\varphi_{{\bf p}_n}}$
corresponds to the right vertex of the diagram in
Fig.~\ref{impureCooperloop}.

In order to sum up the whole Cooper ladder,
it is useful to employ a recurrent relation between
the ladder diagrams of the $n$-th and the $n+1$-t order:
\begin{eqnarray}\label{LLeq}
&&L^{n+1}_{\lambda_{n+1}}(\omega,\varphi_{{\bf p}_{n+1}})=
\sum_{\lambda_{n}}\int \frac{d\varphi_{{\bf p}_n}}{2\pi}
L^{n}_{\lambda_n}(\omega,\varphi_{{\bf p}_{n}}) \times \nonumber\\
&&\times C_{\lambda_n}(\omega,\sin{\varphi_{{\bf p}_n}})
V_{\lambda_n\lambda_{n+1}}
(\varphi_{{\bf p}_n}, \varphi_{{\bf p}_{n+1}}).
\end{eqnarray}
The form of Eq.~(\ref{amplitude}) helps to identify an Anzats
\begin{eqnarray}\label{Anzats}
\!L^n_{\lambda_n}(\omega,\varphi_{{\bf p}_{n}})\!=\!
\{l_n^0(\lambda_n,\omega)+
l_n^1(\lambda_n,\omega)
\sin{\varphi_{{\bf p}_n}}\}e^{-i\varphi_{{\bf p}_n}}
\end{eqnarray}
for the solution which is consistent with the recurrent
relation~(\ref{LLeq}). After substituting the
Anzats~(\ref{Anzats}) in Eq.~(\ref{LLeq}), we encounter integrals
\begin{eqnarray}\label{integrals}
I_{\lambda}^j=\frac{1}{4\tau}\int_0^{2\pi}
\frac{d\varphi}{2\pi}\frac{i\sin^j\varphi}{i\bar\omega
-H_{\lambda}\sin{\varphi}}, \quad \text{where   } j=0,1,2.
\end{eqnarray}
Then equation~(\ref{LLeq}) can be rewritten in the matrix form
\begin{equation}
\vec{l}_{n+1} = \hat{R}\vec{l}_n,
\label{R}
\end{equation}
where we define a 4-vector
\begin{eqnarray}\label{4vector}
\vec{l}_n^T =
(l^0_{n}(+)\, , l^1_{n}(+) \, , l^0_{n}(-)\, , l^1_{n}(-))
\end{eqnarray}
and a $4\times 4$ matrix
\begin{eqnarray}\label{44matrix}
\hat{R} =
\left(\!\begin{array}{cccc}
I^0_{+} & I^1_{+} & -I^0_{-} & -I^1_{-}\\
I^1_{+} & I^2_{+} & I^1_{-} & I^2_{-} \\
-I^0_{+} & -I^1_{+} & I^0_{-} & I^1_{-}\\
I^1_{+} & I^2_{+} & I^1_{-} & I^2_{-} \end{array} \!\right),
\end{eqnarray}
with the three ``block integrals'', Eq.~(\ref{integrals}),
calculated as
\begin{eqnarray}
I_{\lambda}^0=\frac{1}{\sqrt{\bar\omega^2+
H_{\lambda}^2 }}\frac{1}{4\tau},
\nonumber \\
\nonumber \\
I_{\lambda}^1=
\frac{i}{H_{\lambda}}\left(\frac{|\bar\omega|}{\sqrt{
\bar\omega^2+H_{\lambda}^2 }}-1\right)\frac{1}{4\tau},
\nonumber \\
\nonumber \\
I_{\lambda}^2=
-\frac{|\bar\omega|}{h^2_{\lambda}}\left(\frac{|\bar\omega|}
{\sqrt{\bar\omega^2+H_{\lambda}^2 }}-1\right)\frac{1}{4\tau}.
\end{eqnarray}

Now we can proceed with the calculation of the Cooper ladder.
We substitute the Anzats~(\ref{Anzats}) and the
Cooper block~(\ref{Cooper_block})
in the Cooper ladder~(\ref{C}), and obtain
\begin{eqnarray}\label{scalarproduct}
{\cal{C}}&=&4\pi\tau\nu(\epsilon_F)T\sum_{\omega>0}
\sum_{n=0}^{\infty}\sum_{\lambda}\lambda
\left(l_n^0(\lambda)I_{\lambda}^0+l_n^1(\lambda)I_{\lambda}^1\right)=
\nonumber \\
&=&4\pi\tau\nu(\epsilon_F)T
\sum_{\omega>0}\vec{I}^T\cdot(1-\hat{R})^{-1}\cdot
\vec{l}_0,
\end{eqnarray}
where we used the definition of the integrals~(\ref{integrals})
and the 4-vectors~(\ref{4vector}), and we summed up the geometric
progression $\sum_{n=0}^{\infty}\hat{R}^n=(1-\hat{R})^{-1}$. We
also introduced a vector $\vec{I}^T=(I_+^0,I_+^1,-I_+^0,-I_+^1)$
and used a notation $\vec{l}_0^T=(1,\;0,\;-1,\;0)$, corresponding
to the definition of $L^0_{\lambda_0}(\omega,\varphi_{{\bf
p}_{0}})$. At this point we have as well neglected the difference
$\nu_+\neq\nu_-$.

Evaluating the scalar product in Eq.~(\ref{scalarproduct}),
we find  the Cooper ladder ${\cal C}$  and  the equation for $T_c$:
\begin{equation}\label{ImTc}
\frac{1}{\nu(\epsilon_F)U(0)}=\pi T\max_{q} \sum_{\omega>0}
K\left(\omega,H_+,H_-,\frac{1}{2\tau}\right),
\end{equation}
where the kernel is
\begin{equation}\label{bigimpuritykernel}
K(\omega) =4\tau \frac{(I^0_++I^0_-)\left[1-(I^2_++I^2_-)\right]+
(I^1_+-I^1_-)^2}{\left(1-(I^0_++I^0_-)\right)
\left[1-(I^2_++I^2_-)\right]-(I^1_+-I^1_-)^2}.
\end{equation}
Equation (\ref{ImTc}) for $T_c(h)$ was solved numerically; the
phase transition lines are shown in Fig.~\ref{tau} for a number of
impurity scattering strengths $1/2\tau T_{c0}$, interpolating from
a clean to dirty limit. It is seen that in the clean limit
$T_{c0}\tau \gg 1 $ the impurity scattering decreases the critical
parallel magnetic field (in the clean limit it is given by
$H_{p0}=\sqrt{2\alpha p_F\Delta(0)} $ first found
in~[\onlinecite{BG}]) and simultaneously pushes the position of
the ${\cal L}$ point to higher values of $H$  and lower values of
$T$. As a result, both short-wavelength inhomogeneous states
disappear from the phase diagram at $\tau^{-1} \geq 9 T_{c0}$.
Note a large numerical factor 9 which enters the definition of the
``dirty'' regime in the present problem. In the ``dirty limit''
$1/\tau\geq 10T_{c0}$ the kernel $K(\omega)$ simplifies to
\begin{equation}\label{dirty}
K(\omega)=\frac{2}{|\omega|+2H^2\tau+v_F^2Q^2\tau/4}.
\end{equation}

The kernel (\ref{dirty}) is maximal at $Q=0$, i. e. in the dirty
limit one obtains a homogeneous superconducting phase (see below,
however). The zero-temperature limit of the paramagnetic critical
magnetic field $H_{p}(0)$ can be easily obtained with the use of
Eq.~(\ref{dirty}) as
\begin{equation}
H_{p}=\sqrt{\frac{\pi T_{c0}}{4\tau e^{\gamma}}}.
\end{equation}
Thus, in the dirty limit $1/\tau\gg T_{c0}$ the paramagnetic
critical field {\it grows} with increase of disorder.

The above results in this Section were obtained in the main order
over the parameter $\alpha/v_F$. The linear in $\alpha/v_F$ terms
can be included in the same way it was done in the preceding
Section, which lead to the substitution of $I_\lambda^j
\rightarrow I_\lambda^j(1+\lambda\alpha/v_F)$ in the kernel
(\ref{bigimpuritykernel}). Then in the dirty limit the kernel
reads
\begin{equation}\label{dirtymodified}
K(\omega)=\frac{2}{|\omega|+2H^2\tau+v_F^2Q^2\tau/4+2\tau\alpha QH}.
\end{equation}
Maximization of the Cooper loop with the kernel
(\ref{dirtymodified}) with respect to the pairing vector $Q$
leads to a non-zero momentum of the Cooper pair, equal to
\begin{equation}\label{dirtyqhel}
Q_{hel}=-\frac{4\alpha H}{v_F^2},
\end{equation}
for all values of the magnetic field.
Thus the terms $\alpha/v_F$ transform
the homogeneous superconducting state into a weakly inhomogeneous
helical state, analogously to the clean case.
Note, that the small wave vector modulating the order parameter
in the ``dirty'' limit, Eq.~(\ref{dirtyqhel}),
is twice larger than it is  in the clean case
for weak magnetic fields, Eq.~(\ref{Qhel}).
Note, however, that the result (\ref{dirtyqhel})
was obtained near the transition line $T_c(H)$ only.

\section{Fluctuational effects near the $T_c(H)$ transition line}
\label{BKTtransition}

The corrections to the mean-field approximation, employed in this
paper, are usually of the order of $T_c/\epsilon_F$ for a clean 2D
superconductor: the actual transition is of the
Berezinsky-Kosterlitz-Thouless vortex depairing type, and the
transition temperature is shifted downwards
 by a relative amount  $T_c/\epsilon_F \ll 1$.
In our system the fluctuations are enhanced strongly around the
special points ${\cal L}$ and ${\cal S}$, where an additional
analysis is needed.

We start from the ${\cal L}$ point and recall  that near the whole
${\cal LT}$ line the component $n^{yy}_s$ of the superfluid
density is suppressed, cf. Eq.~(\ref{SCdensity}), proportionally
to the coefficient $a$ from Eq.~(\ref{afirst}). Within the
approximation used in Sec.~\ref{Current} and above, this reduction
factor could be arbitrary small, since  $n^{yy}_s$ vanishes
exactly at the ${\cal LT}$ line. However, as it was explained in
Sec.~\ref{Qsmall}, an account of the sub-leading terms $\propto
\alpha/v_F$ transforms the ${\cal LT}$ line of the Lifshitz phase
transition into a crossover region. The minimal value of the
second derivative $d^2\Omega_{hel}(Q)/dQ^2$ and thus of the ratio
$n^{yy}_s/n^{xx}_s$ then scales as $\eta^{2/3} \propto
(\alpha/v_F)^{2/3}$. In a superconductor with an anisotropic
tensor of the superfluid density, the effective ``rigidity
modulus'' is controlled by the geometric average
$\sqrt{n^{xx}_sn^{yy}_s}$. The strength of the phase fluctuations
is thus larger near the ${\cal LT}$ line by a factor
$(v_F/\alpha)^{1/3}$. Therefore, the fluctuational reduction of
the transition temperature near the ${\cal L}$ point is enhanced
by the same relative factor $(v_F/\alpha)^{1/3} \gg 1$, cf.
Fig.~\ref{phdBKT}.

Another mechanism of a fluctuation enhancement is effective near
the ${\cal S}$ point due to the extended $U(2)$ symmetry of the
order parameter. Exactly at the symmetric point  ${\cal S}$ the
order parameter spinor spans the sphere $S^3$ and is equivalent to
a four component unit vector, see Eq.~(\ref{4N}). The thermal
fluctuations of the classical $O(N)$ nonlinear vector model in 2D
space have been studied by Polyakov~[\onlinecite{PolyakovRG}]. On
a large length scales $L \gg \xi(T)$, where $\xi(T)$ is the
temperature-dependent superconductive correlation length, the
evolution of the dimensionless coupling constant $g =
T\left(\frac{\hbar^2}{2 m}n_s\right)^{-1} $ is governed (for
$N=4$) by the renormalization-group equation
\begin{equation}\label{rngP}
\frac{d g}{d X}= \frac{2}{\pi} g^2,
\label{RGg}
\end{equation}
where $X=\log(L/\xi(T))$. Deviations from the symmetric point are
measured (cf. Sec.~\ref{Current}) by the anisotropy parameter
$\beta_a$ which enters into the free energy in the combination
\begin{equation}
\beta_{a}(T,H) |\Delta|^4({\cal N}_1^2+
{\cal N}_2^2-{\cal N}_3^2-{\cal N}_4^2)^2.
\label{betaaa}
\end{equation}
It is easy to show the following~\cite{PolyakovRG}:
that the anisotropy parameter
$\beta_a$ satisfies a renormalization group equation:
\begin{equation}
\frac{d \beta_a}{d X}=-\frac{4 g}{\pi}\beta_a.
\label{RGbeta}
\end{equation}
Running solution of Eqs.~(\ref{RGg}, \ref{RGbeta}) can be easily
found:
\begin{equation}\label{}
g^{-1} = g_0^{-1} - \frac2{\pi} X
\label{Ns}
\end{equation}
and
\begin{equation}
\beta_a = \beta_a^0 \left(1 - \frac{2g_0}{\pi} X\right)^2
\label{beta_a}
\end{equation}
(a renormalization of the coefficient $\beta_s$ is absent within
the same approximation). The infra-red cutoff length $L_{\rm
max}(T)$ for the renormalization flow solution (\ref{Ns},
\ref{beta_a}) coincides with the length $L_{\rm an}$ defined in
Eq.~(\ref{Lan}). Finally, we obtain the renormalized parameters
\begin{equation}
n_s=n_s^0-\frac{2Tm}
{\pi\hbar^2}\log\left(\frac{\beta_s}{\beta_a}\right) \label{ns2}
\end{equation}
and
\begin{equation}
\beta_a=\beta_a^0\left (1-\frac{2Tm} {\pi\hbar^2 n_s^0}
\log\left(\frac{\beta_s}{\beta_a}\right) \right)^{2}.
\end{equation}
The above formulae are valid as long as the effective
renormalization group ``charge'' $g(X_{\rm max}) $ is small
compared to unity; within this domain a strong suppression of
$\beta_a$  is still possible. Qualitatively, the result of the
$U(2)$-fluctuations is twofold: i) the ``nearly-isotropic''
behavior extends to a wider region around the ${\cal S}$ point,
and ii) the transition temperature occurs to be additionally
suppressed within the same region.

To conclude this Section, fluctuations lead to the deformation of
the phase transition line $T_c(H)$  in the vicinities of the
${\cal L}$ and ${\cal S}$ points, as shown in Fig.~\ref{phdBKT}.

\psfrag{L}[c][c][4][0]{\kern0pt\lower-1pt\hbox{${\cal L}$}}
\psfrag{S}[c][c][4][0]{\kern3pt\lower-2pt\hbox{${\cal S}$}}
\psfrag{T}[c][c][4][0]{\kern1.4pt\lower-2pt\hbox{${\cal T}$}}
\psfrag{T1}[c][c][4][0]{\kern2pt\lower0pt\hbox{${\cal T'}$}}
\begin{figure}
\includegraphics[angle=0,width=0.485\textwidth]{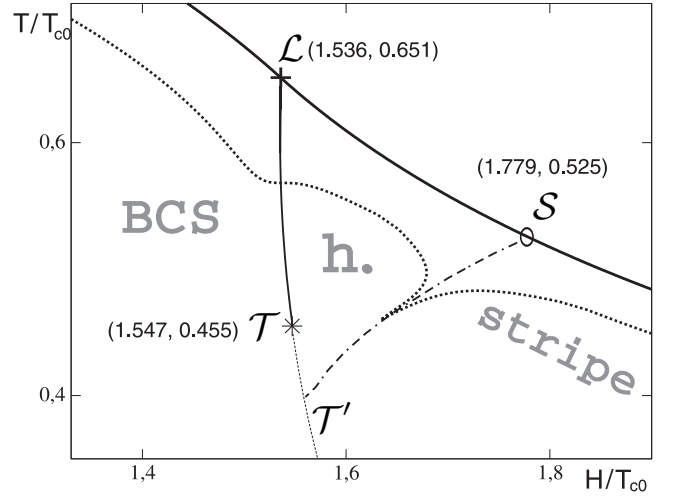}
\caption{\label{phdBKT} The phase diagram with the physical
$T_{BKT}(H)$ transition line shown by a dotted line, for the
values $T_{c0}/\epsilon_F=0.02$ and $\alpha/v_F=0.34$. The effect
of the enhanced fluctuations in the regions near the points ${\cal
L}$ and ${\cal S}$ is seen as a shift of the $T_{BKT}(H)$ line in
the direction of lower temperature and field, with respect to its
mean-field position. The mean-field phase transition line $T_c(H)$
is shown by a solid line. }
\end{figure}

\section{Conclusions}

In this paper we found the phase diagram of a surface
superconductor with a relatively strong Rashba splitting of the
electron spectrum: $T_c \ll \alpha p_F \ll \epsilon_F$, in the
presence of a parallel magnetic field, both in clean and
disordered cases. In the clean limit, we demonstrated that in the
lowest approximation over the spin-orbital parameter $\alpha/v_F$
the phase diagram is universal and contains the unusual
``helical'' phase with the order-parameter modulation $\propto
\exp(i {\bf Q r})$, as well as a usual BCS-type phase and a LOFF
phase with a sinusoidal modulation at higher magnetic fields. The
absolute value of the modulation wavevector is typically $Q \sim
g\mu_B h/\hbar v_F$. Non-magnetic impurities tend to diminish the
region where  these modulated phases exist, and eliminate them
completely in the dirty limit.  Once sub-leading terms of the
order of $\alpha/v_F$ (which are due to the different chiral
subband DoS) are taken into account, the uniform BCS state becomes
unstable and transforms into a ``weakly modulated'' helical phase
with $Q_{\rm slow}\sim \alpha g\mu_B h/\hbar v_F^2$. A weakly
modulated phase is stable with respect to disorder, as its origin
can be found in the symmetry of the problem.~\cite{Agterberg}

In the clean limit, we were able to determine the positions of the
phase transition lines between the BCS, the helical and the LOFF
states on the $(H,T)$ phase diagram. It was found possible to
implement this calculation mainly analytically, due to the fact
that the absolute value of the order parameter is constant in the
helical state. We have shown that the thermal phase fluctuations
are enhanced strongly around both these transition lines, leading
to local ``caves'' of the $T_c(h)$ transition line near the
end-points  ${\cal L}$ and ${\cal S}$. We expect that a
fluctuational correction to the conductivity of the
Aslamazov-Larkin type~\cite{AL1968} should be strongly anisotropic
near the ${\cal L}$ point, which could be one of the  signatures
of the proposed phase diagram.

We have shown that an electric current does not flow in the
equilibrium state of our system at any parallel magnetic field (as
long as the system is infinite), contrary to the statement made
in~[\onlinecite{Yip}].  However, if the system is considered in
the finite-stripe geometry with the periodic boundary conditions,
an oscillating (as a function of the parallel magnetic field)
current is expected, which is another special feature of a
superconductor with a strong Rashba coupling.

A new type of a vortex-like topological defect was predicted to
exist in the region near the special ${\cal S}$ point of the phase
diagram, due to the extended $U(2)$ order parameter symmetry
realized at this point. Contrary to the usual singular vortex,
this new topological defect is non-singular in the sense that the
amplitude of the order parameter does not vanish in the vortex
core. We showed that the energy of the non-singular vortex  is
lower than that of the singular vortex in some finite region
around the ${\cal S}$ point. The nature of an elementary vortex
defect differs between the helical and the stripe phases (in the
regions to the left and to the right from the ${\cal S}$ point):
whereas in the helical state a vortex carries an integer flux
$\Phi = n \Phi_0$ and an integer topological charge $Q = n$, these
quantum numbers are {\it half-integer} in the stripe state.
Physically, these half-vortices correspond to the presence of a
vortex-like defect within only one (either $\Delta_+$ or
$\Delta_-$) component of the order parameter.

{\it Note added.}---When this paper was nearly completed, we
learned about a recent preprint by Agterberg and
Kaur~\cite{Agterberg2}, where the density of states effects (due
to $\nu_+ \neq \nu_-$) upon the phase diagram of the Rashba
superconductor were studied in the clean limit by means of a
numerical solution of the Eilenberger equations.  The main
emphasis was made on the 3-dimensional systems there, thus the
comparison is not straightforward. Qualitatively, the results they
obtained seem to be in agreement with ours.

\section*{Acknowledgments}
We thank  Yu. S. Barash, V. B. Geshkenbein, L. B. Ioffe, A. B.
Kashuba, V. P. Mineev, P. M. Ostrovsky, H. R. Ott, D. A. Ivanov,
M. Sigrist, M. A. Skvortsov, A. M. Tsvelik, K. S. Turitsyn, G. E.
Volovik for many useful discussions. This research was supported
by SCOPES grant, RFBR grants 01-02-17759, 04-02-16348 and
04-02-08159, and Program ``Quantum macrophysics'' of Russian
Academy of Sciences. The research of O. D. was also supported by
the Dynasty Foundation and the Landau Scholarship (FZ-Juelich).

\appendix
\section*{Appendix}

\subsection{Transformation to elliptic integrals}
\label{Gap}

Consider the integral
\begin{equation}\label{odin}
I_1 = \int_{0}^{2\pi}\frac{d\varphi}{2}
\frac{1}{\sqrt{\left(\omega+
iH_{\lambda}\sin\varphi\right)^2+\Delta^2}},
\end{equation}
which enters Eq.~(\ref{stationarygap}),
and note that $\int_0^{2\pi}f(\sin{\varphi})d\varphi=
4\int_{0}^{\pi/2}f(\cos{2\varphi})d\varphi$
for any function $f$.  Using then the identity $\cos{2\varphi} =
\cos^2\varphi - \sin^2\varphi $ and the substitution
$\varphi = \arctan{t}$, we transform Eq.~(\ref{odin})
into
\begin{eqnarray}\label{promeghutok}
I_1 = 2\int_{0}^{\infty}\frac{dt}
{\sqrt{Ae^{i\psi}t^4+
2(\omega^2+H_{\lambda}^2+\Delta^2)t^2+Ae^{-i\psi}}},\nonumber \\
\end{eqnarray}
where we denoted $Ae^{i\psi}=(\omega-iH_{\lambda})^2+\Delta^2$, and
\begin{eqnarray}\label{A}
A=\sqrt{(\omega^2+(|H_{\lambda}|+\Delta)^2)
(\omega^2+(|H_{\lambda}|-\Delta)^2)}.
\end{eqnarray}
The integral (\ref{promeghutok}) is rapidly convergent, thus it is
possible to make a transformation of variables in the complex plane
$\tau=te^{i\frac{\psi}{2}}$ and to reduce Eq.~(\ref{promeghutok}) to
\begin{eqnarray}\label{tri}
I_1 = 2\int_{0}^{\infty}
\frac{d\tau}{\sqrt{A}\sqrt{1+(a+a^{-1})\tau^2+\tau^4}},
\end{eqnarray}
where
\begin{eqnarray}\label{a}
a=\sqrt{\frac{(\omega^2+(|H_{\lambda}|+\Delta)^2)}
{(\omega^2+(|H_{\lambda}|-\Delta)^2)}}.
\end{eqnarray}
(in the course of transformation from Eq.~(\ref{promeghutok}) to
Eq.~(\ref{tri}) we rotated the path of integration over $\tau$ by the
angle $\psi/2$). Making an inverse substitution $\tau = \tan\varphi$
in Eq.~(\ref{tri}) gives
\begin{eqnarray}\label{chetyre}
I_1 = \frac{2}{\sqrt{A}}\int_{0}^{\pi/2}\frac{d\varphi}
{\sqrt{1+\left(\frac{\sqrt{a}-
\sqrt{a^{-1}}}{2}\right)^2\cos^2{2\varphi}}}.
\end{eqnarray}
Following the definition of an elliptic integral of the first kind
${\bf K}(k)=\int_0^{\pi/2}(1-k^2\sin^2{\varphi})^{-1/2}d\varphi$,
we rewrite Eq.~(\ref{chetyre}) as
\begin{eqnarray}
I_1= \frac{2}{\sqrt{A}}{\bf K}
\left(i\frac{\sqrt{a}-\sqrt{a^{-1}}}{2}\right) =
\frac{2}{\sqrt{A}}
\frac{{\bf K}\left(\sqrt{1-\frac{1}{a^2}}\right)}{\sqrt{a}}.
\nonumber \\
\label{fifth}
\end{eqnarray}
The last expression in (\ref{fifth}) leads directly
to the gap equation~(\ref{secline}),
with $A$ and $a$  given by Eqs.~(\ref{A})~and~(\ref{a}).

\subsection{Equation for the $\cal{ST'}$ line}
\label{ST'}
Here we present the transformation
of Eq.~(\ref{eigen}),
which determines
the helical state metastability line $\cal{ST'}$,
to the complete
elliptic integrals of the first and the third kind
(${\bf K}$ and ${\bf \Pi}$):
\begin{eqnarray}
I_2 = &&\int_0^{2\pi}d\varphi\frac{1}
{4\sqrt{\tilde\omega^2+\Delta^2}}
\frac{-1+(\tilde\omega+iX\sin{\varphi})^2}
{\tilde\omega^2+\Delta^2+(X\sin{\varphi})^2}=\nonumber \\
&=&\frac{z{\bf K}(k)-z_1{\bf \Pi}(l_1,k)-z_2{\bf \Pi}(l_2,k)}
{\sqrt{(\Delta+H)^2+\omega^2}}
\label{dvad}
\end{eqnarray}
and
\begin{eqnarray}
I_3 = &&\int_0^{2\pi}\frac{1}{4\sqrt{\tilde\omega^2+\Delta^2}}
\frac{\Delta^2}
{\tilde\omega^2+\Delta^2+(X\sin{\varphi})^2}=\nonumber \\
&=&\frac{y{\bf K}(k)-y_1{\bf \Pi}(l_1,k)-y_2{\bf \Pi}(l_2,k)}
{\sqrt{(\Delta+H)^2+\omega^2}},
\label{trit}
\end{eqnarray}
where $\tilde\omega=\omega+iH\sin\varphi$.
In Eqs.~(\ref{dvad}, \ref{trit}) we introduced
notations
\begin{eqnarray}
l_1({\bf S})&=&1-
\frac{(\Delta-H+i\omega)(\Delta^2-(H+i\omega)^2)+X^2)}
{(\Delta+H+i\omega)
(\Delta^2+H^2+\omega^2-X^2-2i{\bf S})},\nonumber \\
l_2({\bf S})&=&l_1(-{\bf S});\nonumber \\
\mbox{ } \nonumber \\
\mbox{ } \nonumber \\
z&=&-2-\frac{\Delta H
(2i\omega X+\Delta (H+2X))}
{(\Delta +i\omega)^2X^2},\nonumber \\
z_1({\bf S})&=&-\frac{\left(\Delta-H+i\omega\right)
\, L({\bf S})}
{2\left(\Delta+i\omega\right)^2 X^2
\left(\Delta^2-
\left(H-i\omega\right)^2+
X^2\right){\bf S}},\nonumber \\
z_2({\bf S})&=&z_1(-{\bf S});\nonumber \\
\mbox{ } \nonumber \\
y&=&\frac{\Delta^2H^2}{(\Delta+i\omega)^2X^2},\nonumber \\
y_1({\bf S})&=&-\frac{i\Delta^2\left(\Delta-H+i\omega \right)
\,M({\bf S})}
{2\left(\Delta+i\omega\right)^2 X^2
\left(\Delta^2-
\left(H-i\omega\right)^2+
X^2\right){\bf S}},\nonumber \\
y_2({\bf S})&=&y_1(-{\bf S});
\end{eqnarray}
where
\begin{eqnarray}
&&L({\bf S})=-i\Delta^4H(H+X)^2- \nonumber \\
&-&2\omega^2X^2(H-i\omega+X)(\omega X+{\bf S}) +
\nonumber \\
&+&2i\Delta\omega X(H + X)(H-i\omega+X)(\omega X+{\bf S})
+\nonumber \\
&+&\Delta^3
(H+X)^2(-iH^2-H\omega+2\omega X+iX^2+{\bf S})
+ \nonumber \\
&+&\Delta^2(H+X)
(H^2-iH\omega+HX+i\omega X)
(2\omega X+{\bf S}) - \nonumber \\
&-&\omega X^2\Delta^2(H+X)^2, \nonumber \\
\mbox{ }\nonumber \\
&&{\bf S}=\sqrt{-\Delta^2 H^2+\left(\Delta^2+\omega^2\right)X^2},
\quad X=\frac{q+Q}{2};\nonumber \\
\end{eqnarray}
\begin{eqnarray}
M({\bf S})&=&\Delta H^3(\Delta +H-i\omega )- \nonumber \\
&-&HX^2\left(\Delta (\Delta +2H)-i(\Delta -H)\omega +
2\omega ^2\right) +\nonumber \\
&+&(\Delta +i\omega )X^4 + \nonumber \\
&+&\left(H^2(i(\Delta +H)+\omega )+\left(-i(\Delta +H)+
\omega \right)X^2\right){\bf S}.\nonumber \\
\end{eqnarray}

\subsection{The eighth order  terms in the Ginzburg-Landau functional}
\label{8}

Here we present a calculation of the coefficient $\kappa$ which is
in front of the
eighth-order anisotropic term in the free energy,
Eq.~(\ref{Fanis}). As it was mentioned, the sign of $\kappa$
determines the type of the helical state instability.

The terms of the eighth order in $\Delta$ in the
Ginzburg-Landau functional are
\begin{eqnarray}
\label{8order}
&&F_{sn}^{(8)}=
\frac{a}{8}\Delta^8+
\frac{b}{8}(|u|^2-|v|^2)^2\Delta^4+
\nonumber \\
&&\frac{(D_1-D_2+D_3/2)}{8}(|u|^2-|v|^2)^4 +E_{3Q},
\end{eqnarray}
where all but the last term originate from the expansion of the
free energy in powers of the basic amplitudes $u$ and $v$, and
notations $a = D_1+D_2+ D_3/2$ and $b = 6 D_1-D_3$ are introduced.
The last term $E_{3Q}$ arises due to the nonlinearity-induced
additional  harmonics with momenta $\pm 3Q$ and amplitudes $u_{\pm
3Q}$ and $v_{\pm 3Q}$; explicit form of $E_{3Q}$ will be specified
shortly. Actually we are interested in terms which contribute to
the coefficient $\kappa$ in front of the $\cos^4\theta$ term in
Eq.~(\ref{Fanis}), thus  the terms in the first line of
Eq.~(\ref{8order}) are irrelevant.

The coefficients $D_{1,2,3}$ come from the $8$-vertices diagrams
shown in Fig.~\ref{Fig_1}. The corresponding analytical
expressions are summarized below (we introduce notations $D_i =
T\sum_{\omega,{\bf p},\lambda} \tilde{D_i}$):
\begin{widetext}
\begin{eqnarray}
\label{D123} \tilde{D_1}&=&\frac18
G^4_{p+\frac{Q}{2}}G^4_{-p+\frac{Q}{2}}, \qquad\qquad\tilde{D_2} =
G^4_{p+\frac{Q}{2}}G^3_{-p+\frac{Q}{2}} G_{-p-\frac{3Q}{2}}+
G^3_{p+\frac{Q}{2}}G^2_{-p+\frac{Q}{2}}
G^2_{-p-\frac{3Q}{2}}G_{p+\frac{5Q}{2}},  \nonumber \\
\tilde{D_3} &=& \frac32 G^4_{p+\frac{Q}{2}}G^2_{-p+\frac{Q}{2}}
G^2_{-p-\frac{3Q}{2}}+ G^3_{p+\frac{Q}{2}}G_{-p+\frac{Q}{2}}
G^3_{-p-\frac{3Q}{2}}G_{p+\frac{5Q}{2}}+
G^3_{p+\frac{Q}{2}}G^3_{-p+\frac{Q}{2}}
G_{-p-\frac{3Q}{2}}G_{p-\frac{3Q}{2}}+\nonumber \\
&+&
G^2_{p+\frac{Q}{2}}G^2_{-p+\frac{Q}{2}}
G^2_{-p-\frac{3Q}{2}}G_{p-\frac{3Q}{2}}G_{p+\frac{5Q}{2}},
\end{eqnarray}
\end{widetext}
where $G$ are the normal metal Green's functions. The integration
is first done over $d\xi = d(p^2/2m -E_F)$, then over $d\varphi$
by means of a generating function $1/\sqrt{\omega^2+(\lambda H\pm
nQ)^2}$, where $n=1,3,5$. The  obtained analytical expressions are
evaluated for the values $T_S=1.779T_{c0}$, $H_S=0.525T_{c0}$ and
$Q_S=2.647T_{c0}$ at the ${\cal S}$ point of the $T_c(H)$ line.
The summation over $\omega$ is performed numerically, resulting in
\begin{eqnarray}
D_1  =0.00106053, \, \quad D_2=0.00152674, \nonumber \\
D_3=-0.00120067. \label{D123r}
\end{eqnarray}

Now we turn to the evaluation of the ``induced'' term $E_{3Q}$ in
the free energy~(\ref{8order}). It appears due to a generation of
the third harmonics $e^{\pm 3 i Q}$ in the order parameter.  The
amplitudes of the third harmonics $u_{3Q}$ and $v_{3Q}$ are small
and proportional to the third power of the basic amplitudes $u$
and $v$. The ``induced '' term reads
\begin{eqnarray}\label{E3Q}
&&E_{3Q}={\cal V}^+A{\cal V}+(X^+{\cal V}+{\cal V}^+X),
\end{eqnarray}
where the matrix $A$ and the vector $X$ contain numerical
coefficients $l_i$, which are determined by means of evaluation of
the loop diagrams with two, four and six vertices (shown in
Fig.~\ref{Fig_6A}):
\begin{eqnarray}\label{matricaA}
&&A=\left( \begin{array}{cc}
   l_1 +l_3|u|^2+l_{4}|v|^2 & l_{5}uv  \\
    l_{5}u^*v^*&  l_1+l_{4}|u|^2+l_3|v|^2 \\
\end{array} \right),\nonumber \\
\nonumber \\
&&{\cal V}^+=(u_{3Q}^*,v_{-3Q}),\quad X^+=(x^*,y),\nonumber \\
&&x=l_2u^2v^*+l_{6}u^2v^*|u|^2+(l_{7}+l_{8})u^2v^*|v|^2;\nonumber \\
&&y=l_2v^2u^*+l_{6}v^2u^*|v|^2+(l_{7}+l_{8})v^2u^*|u|^2.
\end{eqnarray}
The fifth and higher harmonics, as well as the higher powers of
the third harmonics, do not contribute to the eighth order in
$\Delta$. The amplitudes $u_{3Q}$ and $v_{-3Q}$ in Eq.~(\ref{E3Q})
are found from the minimization of the superconducting energy:
$\frac{\partial E_{3Q}}{\partial {\cal V}}=0$, which gives
\begin{eqnarray}\label{E3Q1}
&&E_{3Q}=-X^+A^{-1}X \equiv\nonumber \\
&&d_2\cdot\left (|u|^6|v|^2+|u|^2|v|^6\right )+
d_3\cdot|u|^4|v|^4
\, ,
\end{eqnarray}
where
\begin{eqnarray}\label{E3Qlast}
&&d_2=
\frac{l_2 (-2 l_{6} l_1 + l_2 l_3)}{l_1^2}\nonumber \\
&&d_3=
\frac{2 l_2 (-2 (l_{7} + l_{8}) l_1 + l_2 (l_{4} + l_{5}))}{l_1^2}.
\nonumber \\
\end{eqnarray}
The analytic expressions for the diagrams shown in
Fig.~\ref{Fig_6A} are (we introduce notations $l_i =
T\sum_{\omega,{\bf p},\lambda} \tilde{l_i}$)
\begin{eqnarray}
&&l_1=\frac{1}{U}-\frac{T}{2}\sum_{\omega,{\bf p},\lambda}
G_{p+\frac{3Q}{2}}G_{-p+\frac{3Q}{2}},\nonumber \\
&&\tilde{l_2}=\frac{1}{2}
G_{p+\frac{Q}{2}}G_{-p+\frac{Q}{2}}
G_{-p-\frac{3Q}{2}}G_{p+\frac{5Q}{2}},\nonumber \\
&&\tilde{l_3}=
G_{-p+\frac{Q}{2}}
G^2_{p-\frac{3Q}{2}}G_{-p-\frac{3Q}{2}},\nonumber \\
&&\tilde{l_4}=G_{-p+\frac{5Q}{2}}
G^2_{p-\frac{3Q}{2}}G_{-p-\frac{3Q}{2}},\nonumber \\
&&\tilde{l_5}=
G_{-p+\frac{Q}{2}}G_{p-\frac{3Q}{2}}
G_{-p-\frac{3Q}{2}}G_{p+\frac{5Q}{2}},\nonumber \\
&&\tilde{l_6}=
G_{p+\frac{Q}{2}}G^2_{-p+\frac{Q}{2}}
G^2_{p-\frac{3Q}{2}}G_{-p-\frac{3Q}{2}},\nonumber \\
&&\tilde{l_7}=\frac{1}{2}
G^2_{p+\frac{Q}{2}}G^2_{-p+\frac{Q}{2}}
G_{p-\frac{3Q}{2}}G_{-p-\frac{3Q}{2}},\nonumber \\
&&\tilde{l_8}=
G_{p+\frac{5Q}{2}}G_{p+\frac{Q}{2}}
G_{-p+\frac{Q}{2}}G_{p-\frac{3Q}{2}}
G^2_{-p-\frac{3Q}{2}}.\nonumber \\
\end{eqnarray}

The calculation of the above integrals in the symmetric point
${\cal S}$ leads to numerical values
\begin{eqnarray}
l_1=0.674254,\qquad l_2=0.103447,\nonumber \\
l_3=0.191039,\qquad l_{4}=0.170477, \nonumber \\
l_{5}=0.206894,\qquad l_{6}=0.0325726, \nonumber \\
l_{7}=0.00491175,\qquad l_{8}=0.021198,
\label{lii}
\end{eqnarray}
which we substitute in Eq.~(\ref{E3Qlast}).
Finally for the coefficient $\kappa$ we obtain a positive value:
\begin{equation}
\kappa = D_1 - D_2 +D_3/2 -d_2 +d_3/2 = 0.0053.
\label{kappaf}
\end{equation}

\begin{figure*}
\includegraphics[angle=0,width=0.43\textwidth]{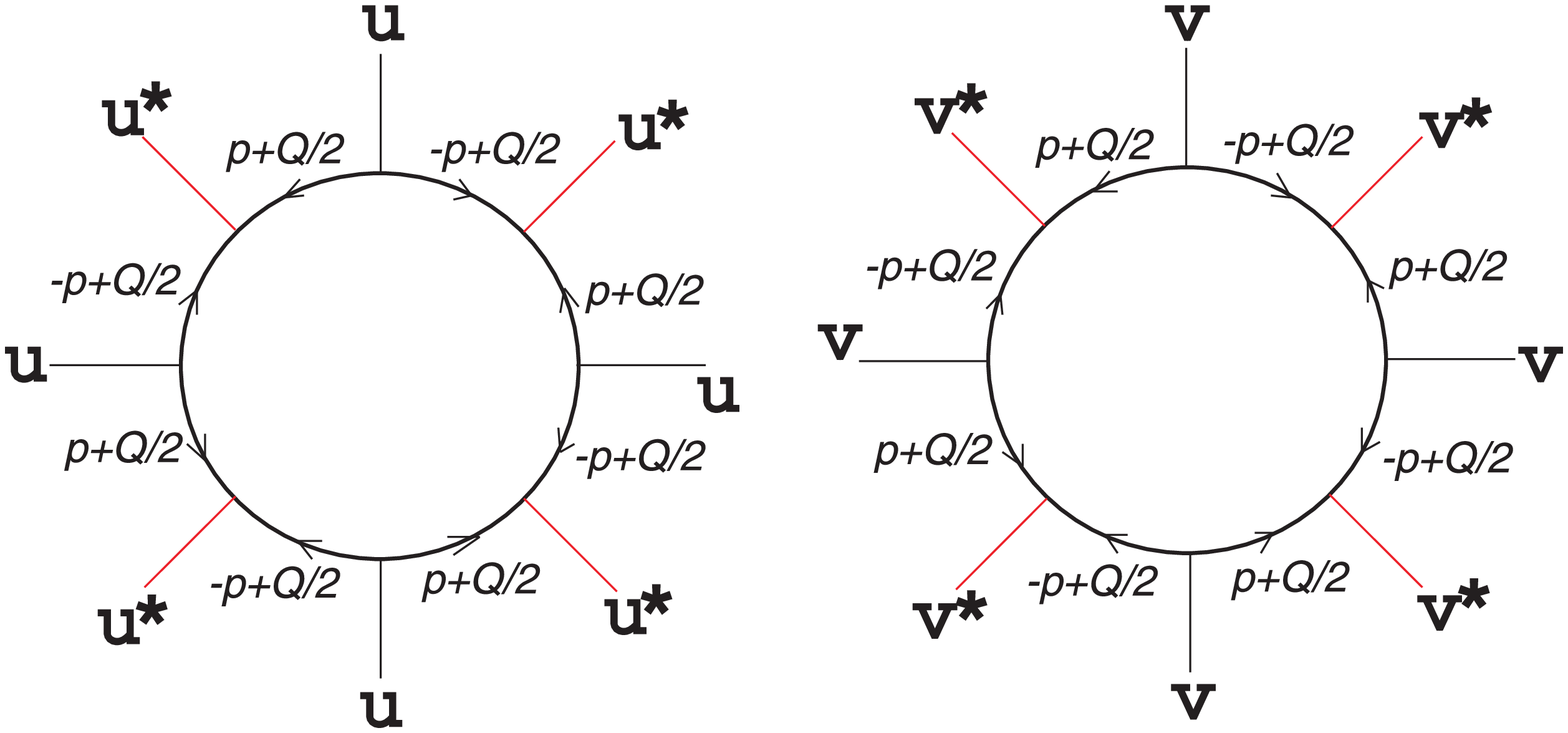}
{\bf a)}
\qquad
\includegraphics[angle=0,width=0.43\textwidth]{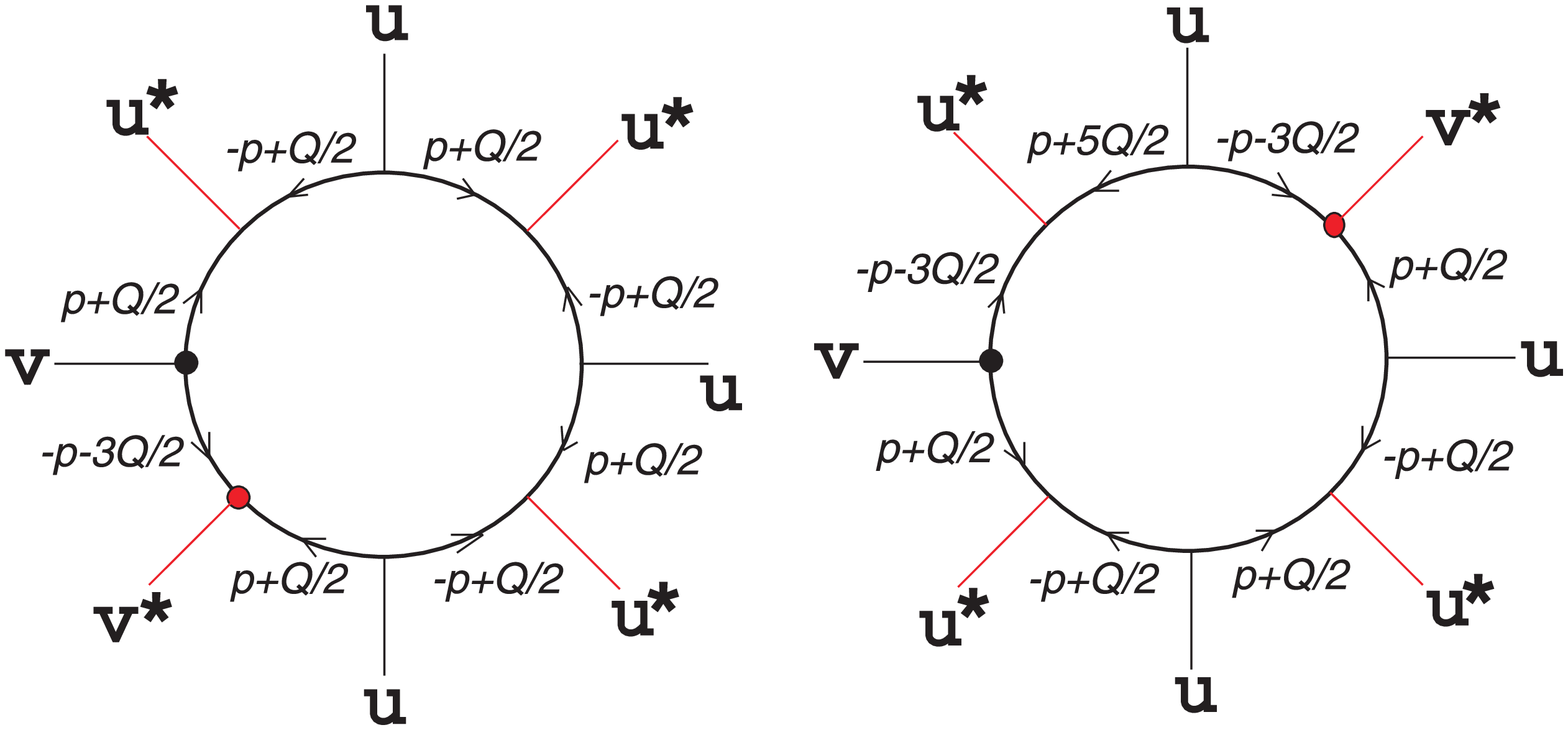}
{\bf b)} \vskip0.7in
\includegraphics[angle=0,width=0.43\textwidth]{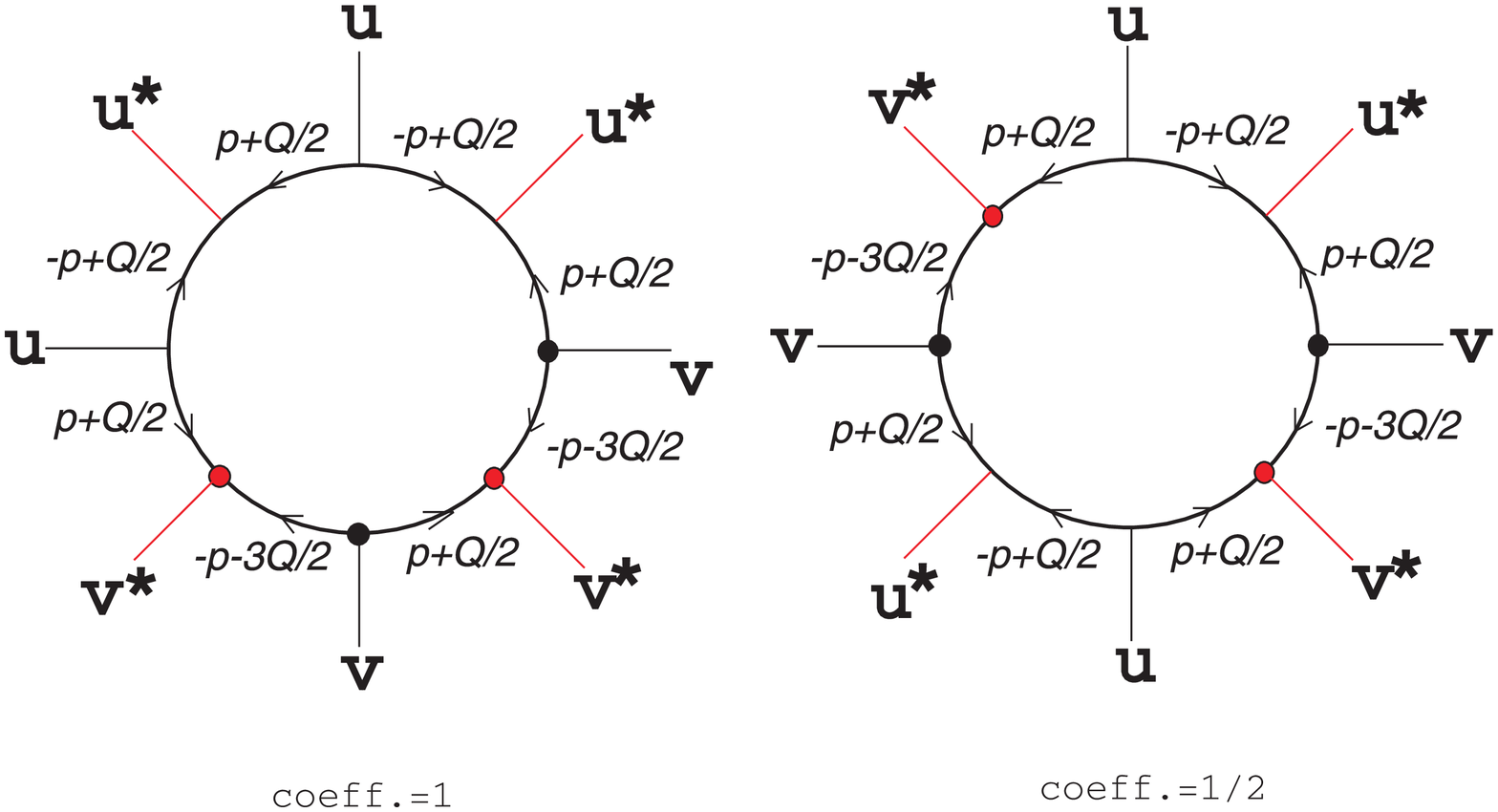}
{\bf c)} \qquad
\includegraphics[angle=0,width=0.43\textwidth]{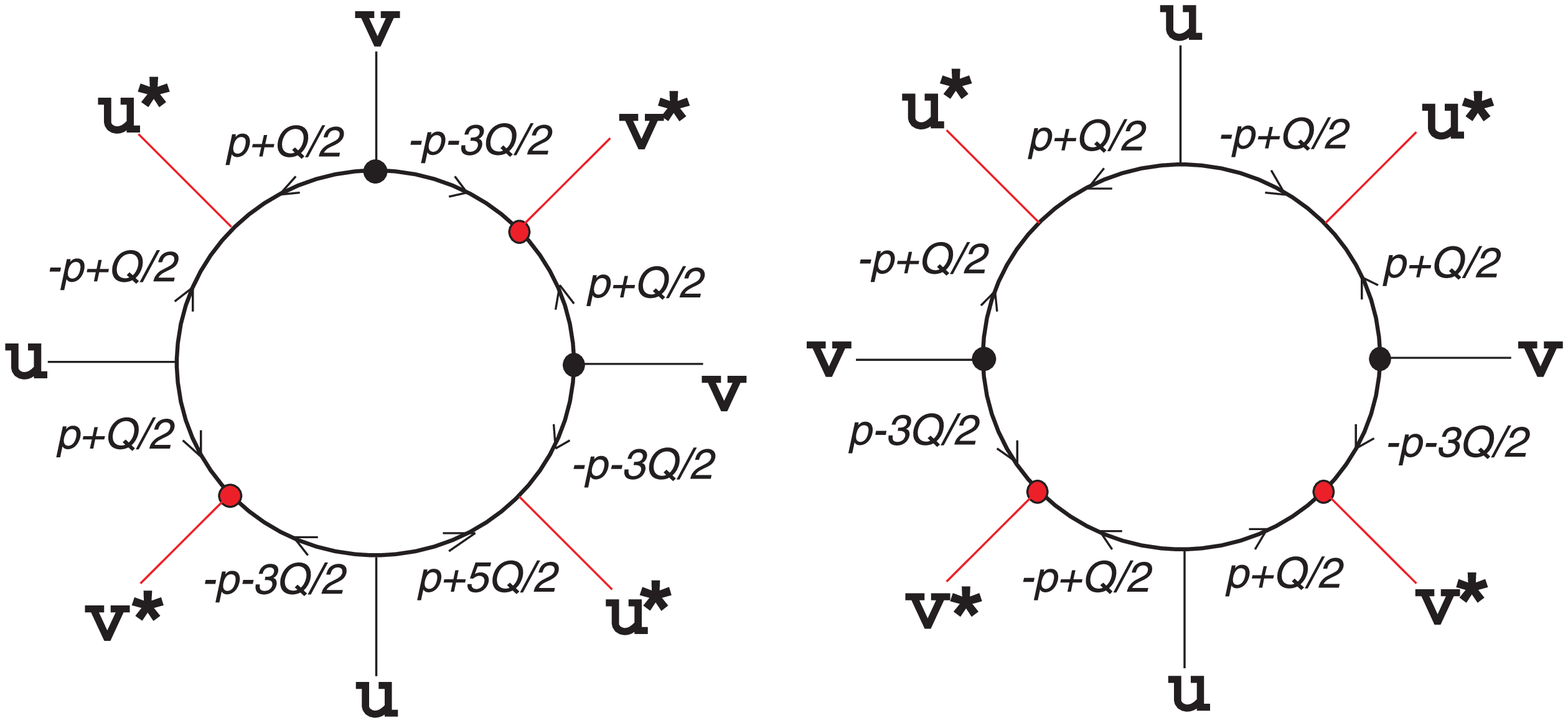}
{\bf d)} \vskip0.7in
\includegraphics[angle=0,width=0.225\textwidth]{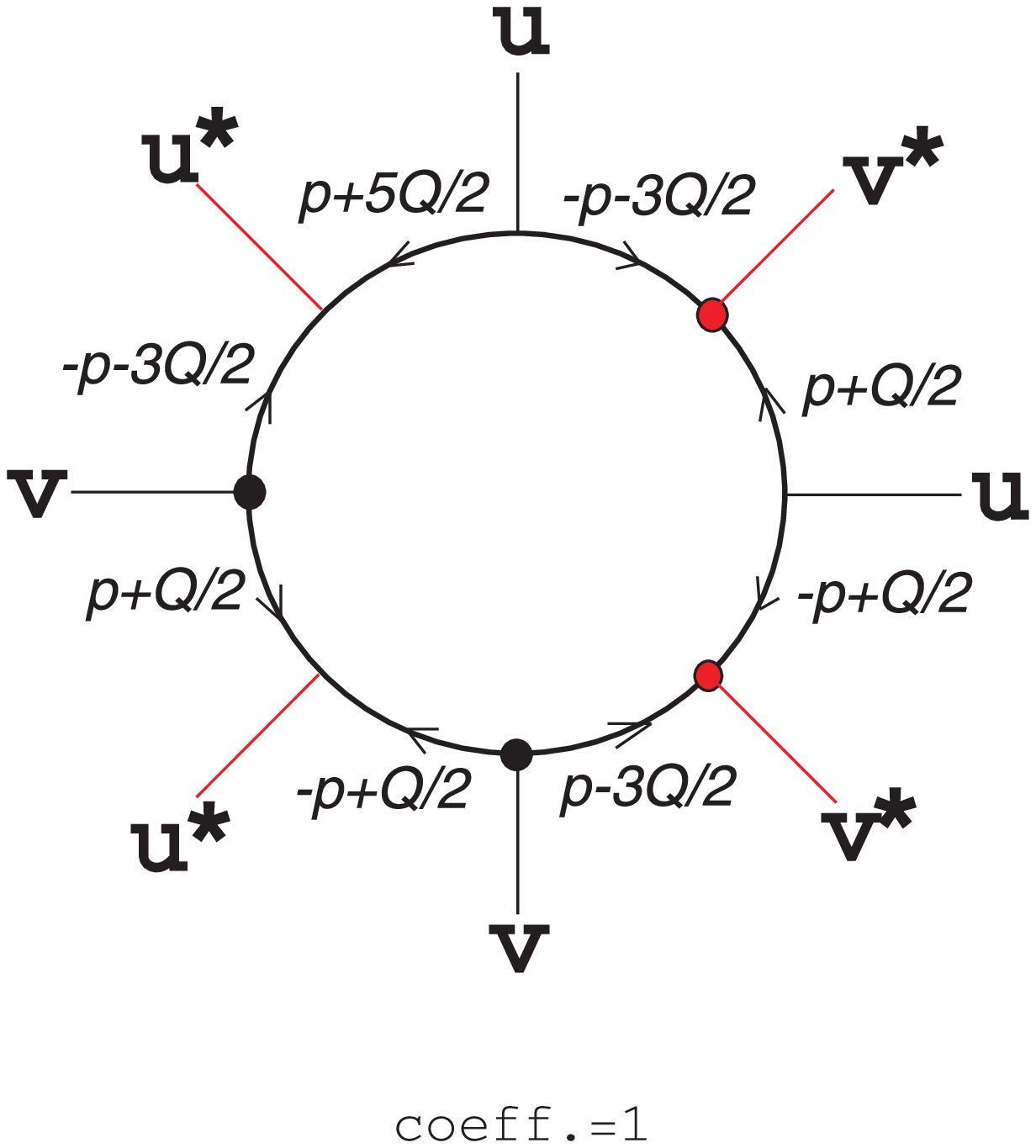}
{\bf e)}
\caption{\label{Fig_1}
\newline
{\bf a)}
Diagrams, corresponding to the coefficient $D_1$ in front
of the term $|u|^8+|v|^8$ in the Ginzburg-Landau expansion. The
combinatorial coefficient of the diagrams is 1/8. All the loops
are symmetric under the replacement $Q\rightarrow-Q$, which is
equivalent to the replacement $u\rightarrow v$. Therefore all such
diagrams are equal.
\newline
{\bf b)}
Two different diagrams, which form the coefficient $D_2$
in front of the term $|v|^2|u|^6$. The combinatorial coefficient
of both is equal to 1.
\newline
The diagrams {\bf c)}, {\bf d)}, {\bf e)}
form the coefficient
$D_3$ in front of the term $|u|^4|v|^4$:
\newline
{\bf c)}
two loops with equal values but different combinatorial
coefficients 1 and 1/2 correspondingly. Under the replacement
$u\rightarrow v$ both diagrams do not change. The complex
conjugate of them simply reverse the encircling along the loop and
such diagrams are taken into account in the combinatorial
coefficients.
\newline
{\bf d)}
the combinatorial coefficient of these both two diagrams
is equal to 1, if we keep in mind the complex conjugate of them.
The two loops transfer one into another under the replacement
$u\rightarrow v$.
\newline
{\bf e)}
a diagram with a combinatorial coefficient equal to 1.}
\end{figure*}

\begin{figure*}
\includegraphics[angle=0,width=0.38\textwidth]{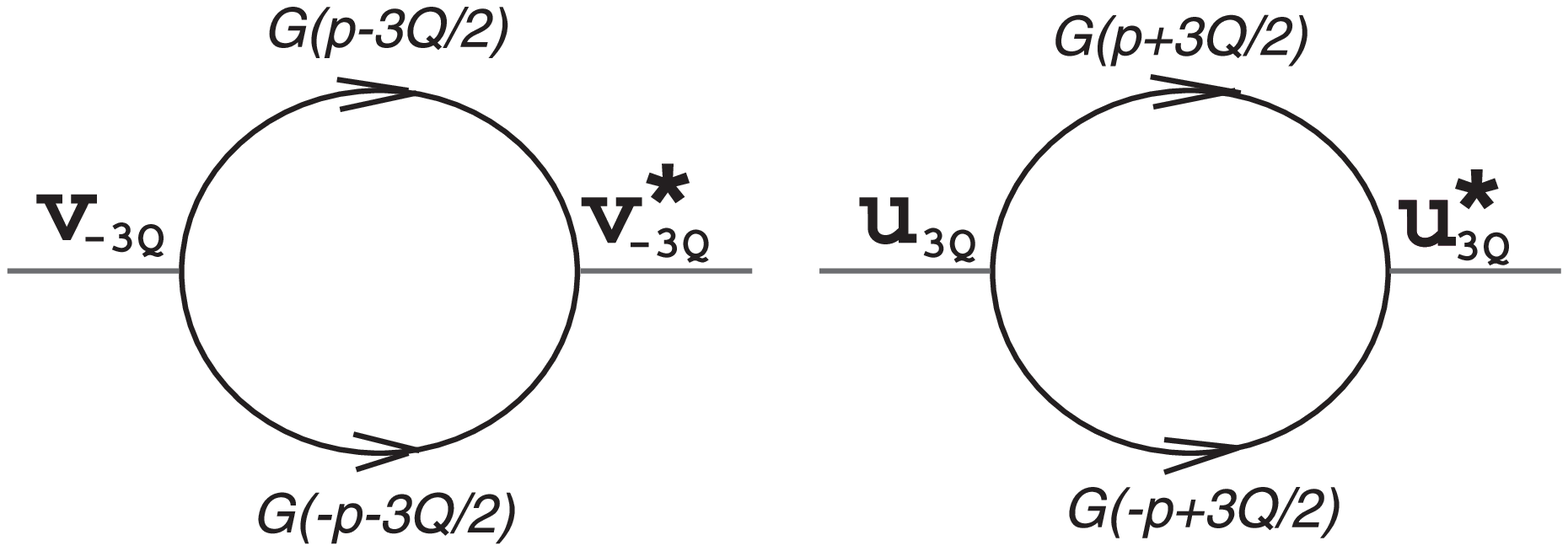}
{\bf a)} \qquad\qquad
\includegraphics[angle=0,width=0.45\textwidth]{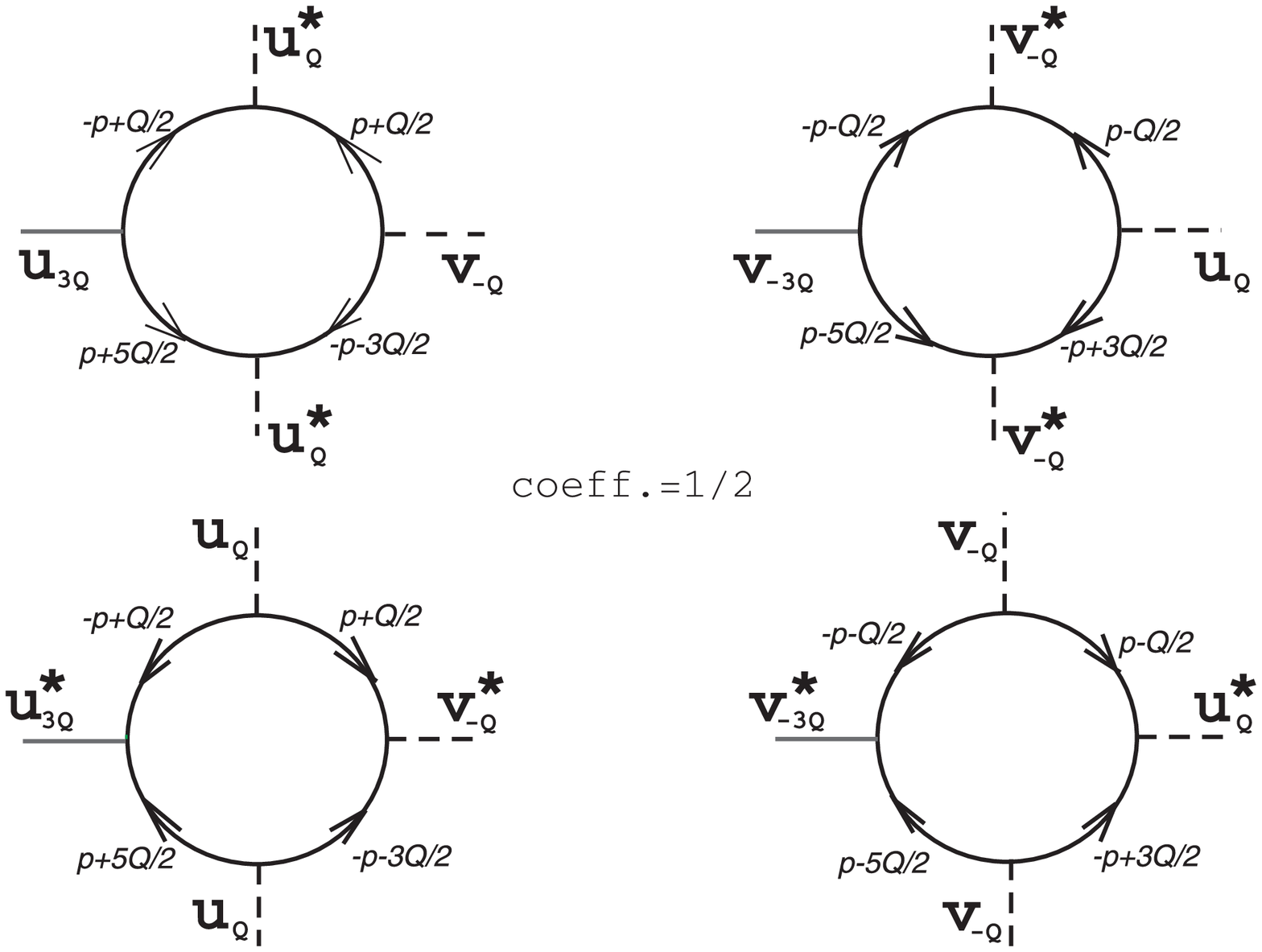}
{\bf b)} \vskip0.5in
\includegraphics[angle=0,width=0.45\textwidth]{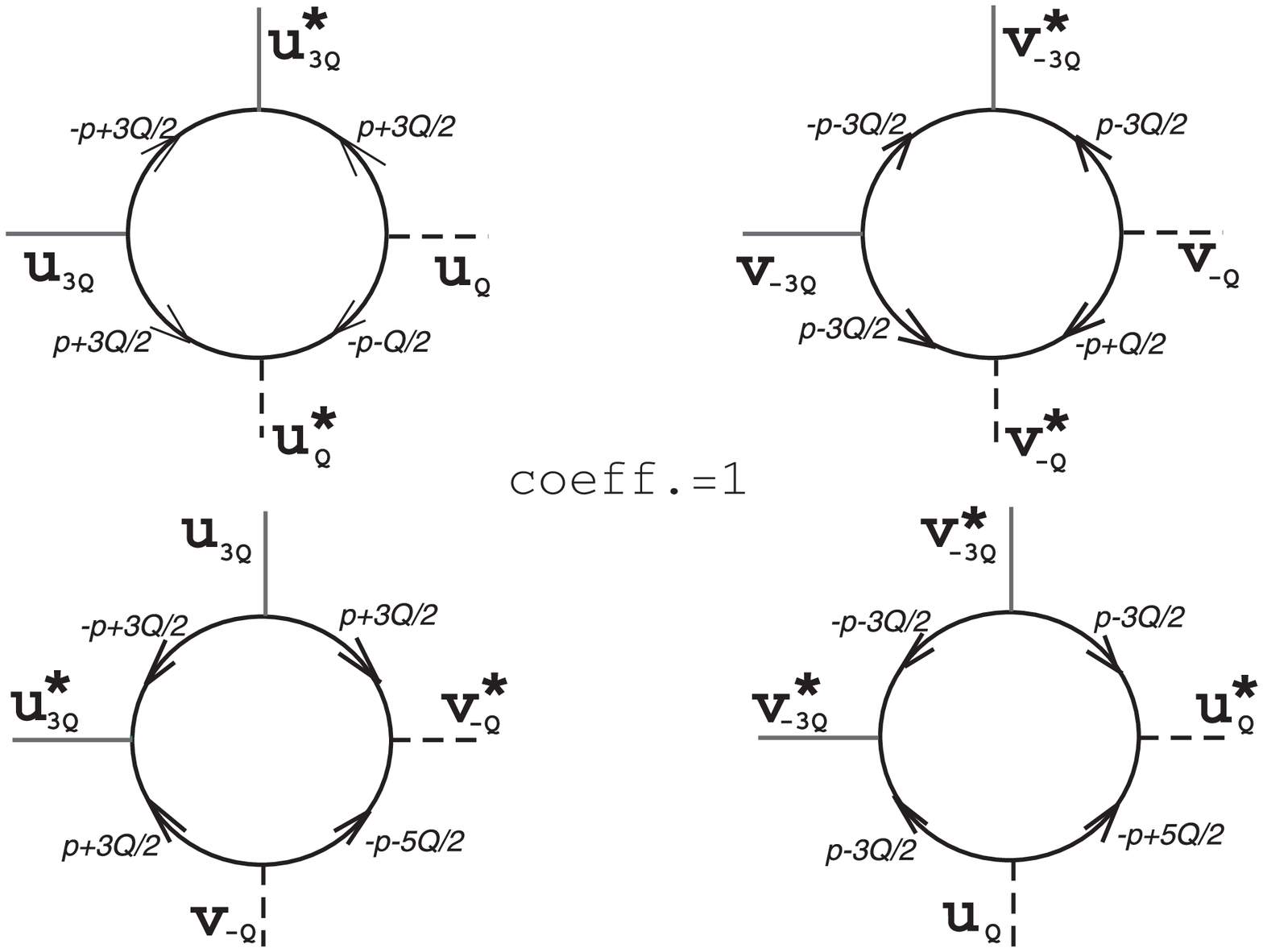}
{\bf c)}\qquad\qquad
\includegraphics[angle=0,width=0.38\textwidth]{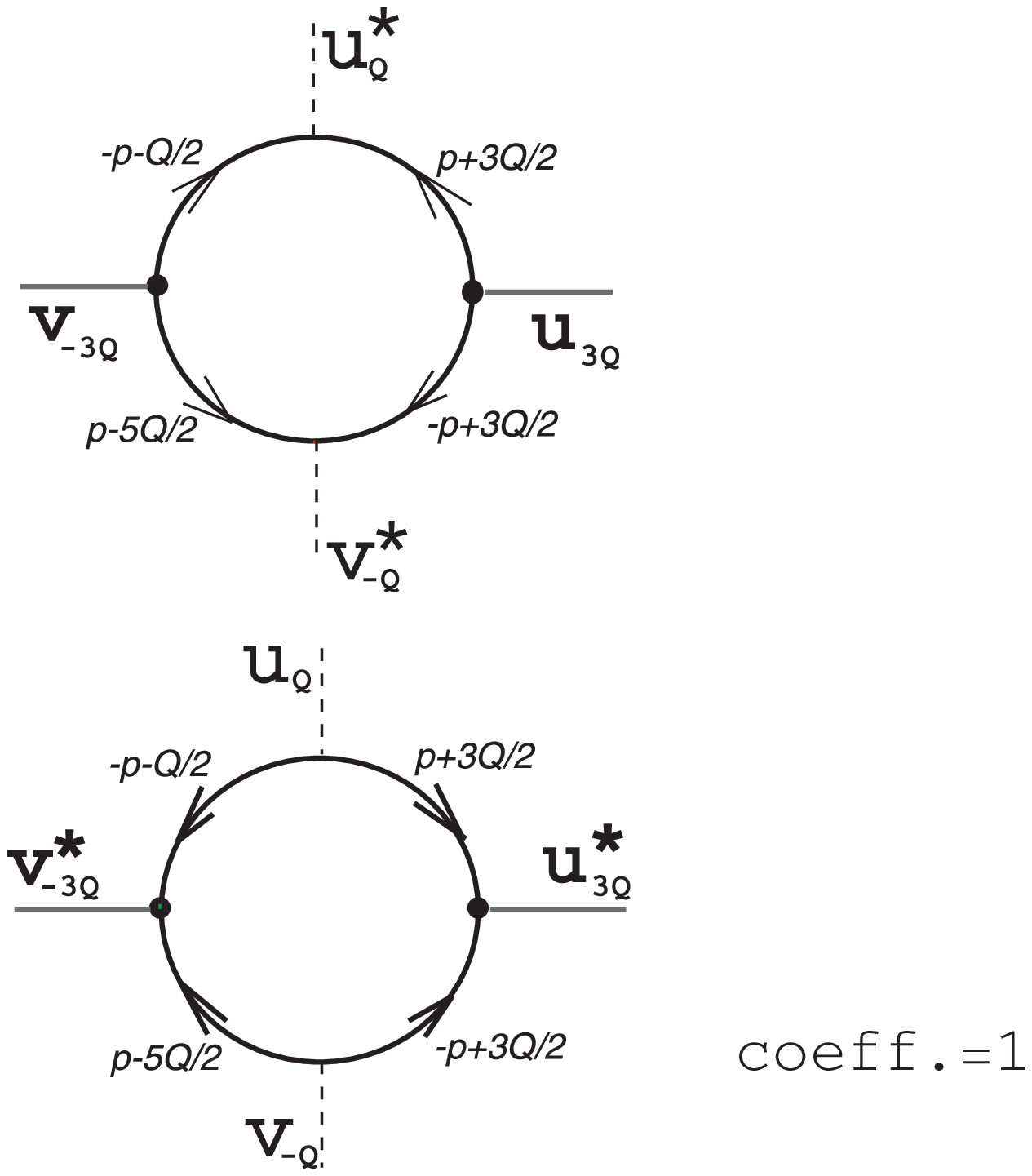}
{\bf d)} \vskip0.5in
\includegraphics[angle=0,width=0.43\textwidth]{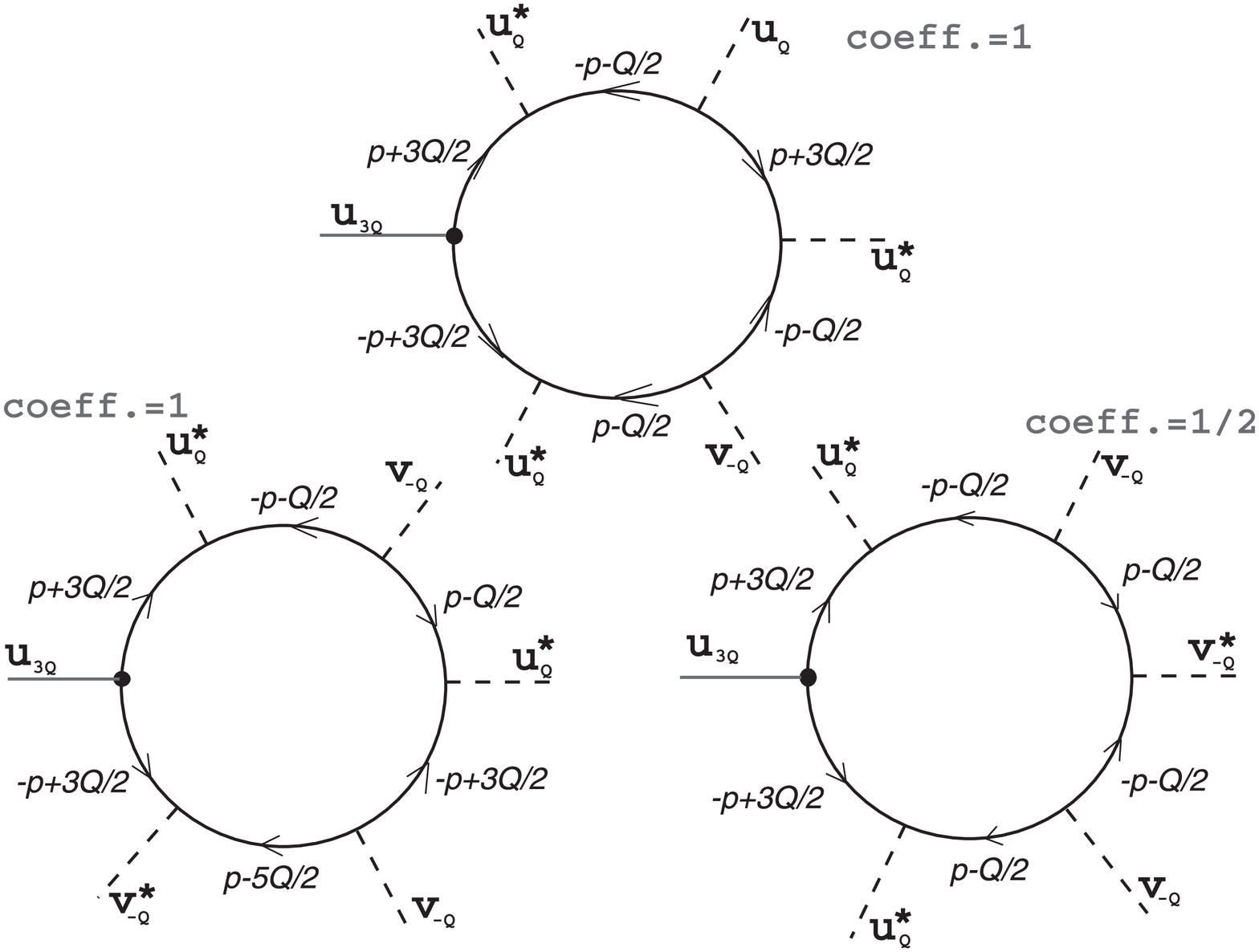}
{\bf e)}
\caption{\label{Fig_6A} Diagrams {\bf a)},
{\bf b)} are of the sixth order in $u$:
\newline
{\bf a)} the  Cooper loop for the third harmonic, which
corresponds to the coefficient $l_1$ in the GL expansion; it has a
combinatorial coefficient equal to 1/2.
\newline
{\bf b)}
diagrams, corresponding to the term
$l_2(u_{3Q}{u_Q^*}^2v_{-Q}+v_{-3Q}{v_{-Q}^*}^2u_Q+h.c.)$. The
combinatorial coefficient of all four is equal to 1/2.
\newline
Diagrams {\bf c)}, {\bf d)}, {\bf e)}
are of the eighth order in
$u$:
\newline
{\bf c)}
diagrams with a combinatorial coefficient of all four of
them equal to 1. The upper two correspond to the term
$l_3(|u_{3Q}|^2|u_Q|^2+|v_{-3Q}|^2|v_{-Q}|^2)$, and the lower two
correspond to the term
$l_{4}(|u_{3Q}|^2|v_{-Q}|^2+|v_{-3Q}|^2|u_Q|^2)$.
\newline
{\bf d)} diagrams, corresponding to the term
$l_{5}(u_{3Q}v_{-3Q}u_Q^*v_{-Q}^*+h.c.)$. The lower loop is a
complex conjugate of the upper loop. The combinatorial coefficient
is equal to 1.
\newline
{\bf e)}
diagrams, corresponding to the terms
$l_{6}u_{3Q}{u_Q^*}^2v_{-Q}|u_Q|^2$ and
$(l_{7}+l_{8})u_{3Q}{u_Q^*}^2v_{-Q}|v_{-Q}|^2$. There are even
more loops which are complex conjugate to the ones shown on the
picture, as well as loops obtained under the replacement
$u\rightarrow v$. The values of such loops are the same, therefore
we do not show them. }
\end{figure*}

\end{document}